\documentclass{aa}  

\usepackage{graphicx}
\usepackage{txfonts}
\usepackage{mathrsfs} 
\usepackage{soul}
\usepackage{siunitx}

\usepackage{color} 

\newcommand{\Msol}{M_{\odot}}

\newcommand{\vv}{\vec{v}}

\newcommand{\vx}{\vec{x}}

\newcommand{\vn}{\vec{n}}

\begin{document} 

   \title{An improved test of the strong equivalence principle with the pulsar in a triple star system}


   \author{G. Voisin\inst{1}\fnmsep\inst{2}\fnmsep\thanks{Email: guillaume.voisin@manchester.ac.uk or astro.guillaume.voisin@gmail.com}
          \and
          I. Cognard\inst{3}\fnmsep\inst{4}
          \and 
          P. C. C. Freire\inst{5}
          \and 
          N. Wex\inst{5}
          \and 
          L. Guillemot\inst{3}\fnmsep\inst{4}
          \and
          G. Desvignes\inst{6}\fnmsep\inst{5}
          \and
          M. Kramer\inst{5}\fnmsep\inst{1}
          \and
          G. Theureau\inst{2}\fnmsep\inst{3}\fnmsep\inst{4}
          }

\institute{Jodrell Bank Centre for Astrophysics, The University of Manchester, Manchester, UK\\
         \and
             LUTH, Observatoire de Paris, PSL Research University, Meudon, France\\
        \and 
            Station de radioastronomie de Nan\c{c}ay, Observatoire de Paris, CNRS/INSU, Universit\'{e} d'Orl\'eans, 18330 Nan\c{c}ay, France \\
        \and
            Laboratoire de Physique et Chimie de l\'{}Environnement, CNRS, 3A Avenue de la Recherche Scientifique, 45071 Orl\'eans Cedex 2, France\\
        \and
            Max-Planck-Institut f\"{u}r Radioastronomie, Auf dem H\"{u}gel 69, D-53121 Bonn, Germany\\
        \and
            LESIA, Observatoire de Paris, Universit\'e PSL, CNRS, Sorbonne Universit\'e, Universit\'e de Paris, 5 place Jules Janssen, 92195 Meudon, France
    }
   
   \date{Received XXXX XX, 2020; accepted XXXX XX, 2020}

 
  \abstract
   {The  gravitational strong equivalence principle (SEP) is a cornerstone of the general theory of relativity (GR). Hence, testing the validity of SEP is of great importance when confronting GR, or its alternatives, with experimental data. Pulsars that are orbited by white dwarf companions provide an excellent laboratory, where the extreme difference in binding energy  between neutron stars and white dwarfs allows for precision tests of the SEP via the technique of radio pulsar timing. }
    { To date, the best limit on the validity of SEP  under strong-field conditions was obtained  with a unique pulsar in a triple stellar system, PSR J0337+1715. We report here on an improvement of this test using an independent data set  acquired over a period of 6 years with the Nan\c cay radio telescope (NRT). The improvements arise from a uniformly sampled data set, a theoretical analysis, and a treatment that fixes some short-comings in the previously published results, leading to better precision and reliability of the test.}
     {In contrast to the previously published test, we use a different long-term timing data set, developed  a new timing model and an independent numerical integration of the motion of the system, and determined the masses and orbital parameters with a different methodology that treats the  parameter $\Delta$, describing a possible strong-field SEP violation, identically to all other parameters. }
{We obtain a violation parameter $\Delta = (+0.5 \pm 1.8) \times 10^{-6}$ at 95\% confidence level, which is compatible with and improves upon the previous study by 30\%. This result is statistics-limited and avoids limitation by systematics as previously encountered. We find evidence for red noise in the pulsar spin frequency, which is responsible for up to 10\% of the reported uncertainty. We use the improved limit on SEP violation to place constraints on a class of well-studied scalar-tensor theories, in particular we find $\omega_{\rm BD} > 140\,000$ for the Brans-Dicke parameter. The conservative limits presented here fully take into account current uncertainties in the equation for state of neutron-star matter.}
   {}


   \maketitle
%





\section{\label{sec:intro}Introduction}



Among the fundamental interactions of nature, gravity is unique in attracting all material objects with the same acceleration, at least within current observational precision. This feature of gravity (the universality of free fall, UFF below) was thought by Newton to be a cornerstone of Newtonian mechanics \citep{principia}. Indeed, in the Newtonian theory of gravity, this universal acceleration implies that the inertial mass of a body is always in a fixed proportion to its passive gravitational mass, and is independent of the mass, chemical composition, or the detailed internal structure of the gravitating object. This was presented as an observed physical principle, without a deeper explanation. Newton and many later experimentalists have conducted different experiments to verify UFF, no deviations have been found that are larger than $1.3 \, \times \, 10^{-14}$ \citep{MICROSCOPE}. This equivalence between the inertial and passive gravitational masses for test particles (defined here as objects with negligible gravitational self-energy) is the so-called weak equivalence
principle (WEP).

When thinking about a new theory of gravity that incorporates the laws of special relativity (SR), Einstein had the insight that the gravitational field appears to be absent for a freely falling observer. This was later described by Einstein as the `most fortunate thought in my life' \citep{renn2007genesis}. This idea, that gravity is equivalent to acceleration, naturally explains the WEP. If the relativity principle applies to this situation, then any observers in a sufficiently small room in a free-falling reference frame are not only unable to determine whether the room is in motion or at rest relative to distant bodies, but they are neither able to determine its rate of acceleration in the gravitational field.
This implies that, in the vicinity of the observer the laws of physics are (in very good approximation) given by SR, which means that the Lorentz invariance of SR is obeyed locally (this is the local Lorentz invariance, LLI) and furthermore, that it does not matter where or when an experiment is made (this is known as local position invariance, LPI). The combination of the WEP with LLI and LPI is now known as the Einstein equivalence principle (EEP, \citealt{Will_book}). Schiff's conjecture states that the WEP implies the full EEP for any consistent theory of gravity, for which a strong plausibility argument can be made \citep[see e.g.][]{Will_book}. 

This generalisation of the relativity principle to reference frames in free fall guided Einstein towards general relativity (GR, \citealt{GR}). GR and other metric theories of gravity fulfil the EEP in a natural way: in these theories the gravitational attraction is seen as a result of spacetime curvature, which itself originates from the energy, stress and momentum of the masses in a system, determined by the field equations of the theory. This curvature changes the trajectories of test particles moving within the spacetime (their  `geodesics') in a unique way that does not depend of the detailed nature of the particles themselves, hence the validity of the WEP. Furthermore, for spatial scales that are small compared to the radius of curvature, the geometry of spacetime necessarily approximates the `flat' Minkowski geometry, hence the LLI and LPI automatically apply to non-gravitational experiments. To rephrase, the EEP is a consequence of a universal coupling between matter and gravity \citep{dam12}.

The qualification of `non-gravitational' is key here. If the EEP can be fully extended to gravitational experiments, like the Cavendish experiment, and to objects with large self-gravitational energy, then we have the strong equivalent principle (SEP). This is a crucially important distinction because, while all metric theories of gravity fulfil the EEP, there are suggestive arguments that GR is the only gravity theory in four spacetime dimensions that fully embodies the SEP \citep{2015AmJPh..83...39D, Will_book}\footnote{Nordstr\"om's conformally-flat scalar theory, which is also a metric theory, also fulfils the SEP, however, this is excluded by Solar System experiments \citep{Deruelle_2011}.}.

Therefore, if we are looking for phenomena beyond GR, a promising avenue would be to look for instances of SEP violation. This has an added advantage: if no SEP violation is found, the results of such an experiment can in principle constrain all alternative theories of gravity.




Just as the EEP consists of the WEP, LLI and LPI, the SEP must additionally include gravitational versions of these. Any violations of the LLI and LPI of the gravitational interaction (e.g., the existence of a preferred frame of reference or the location dependence of gravity) have been strongly constrained using pulsar experiments \citep{Shao_Wex_2012,Shao_2013,Shao_LPI}. In what follows, we focus on the gravitational version of the WEP (GWEP, \citealt{Will_book}), which states that the UFF applies not only to test particles, but also to any objects where the gravitational binding energy is important.

For alternative theories of gravity, the gravitational properties of objects generally depend on their amount of self-gravity. This means that at Newtonian level we have a body-dependent effective gravitational constant, $G_{a b}$, meaning the acceleration of a body $a$ in the gravitational field of a body $b$ is given by
\begin{equation}
   \ddot{\vec{x}}_a = -G_{ab} m_b\,\frac{\vec{r}_{ab}}{\|\vec{r}_{ab}\|^3} 
                      + {\cal O}(c^{-2})\,,
\end{equation}
where $m_b$ denotes the inertial mass of body $b$, $\vec{r}_{ab} \equiv \vec{x}_{a} - \vec{x}_{b}$ their (coordinate) separation, and $c$ is the speed of light. $G_{ab}$ depends on the properties of body $a$ and $b$. In the weak-field limit this can be interpreted, to a good approximation, as a mismatch between the inertial and the gravitational masses of the objects:
\begin{equation} \label{eq:Gab}
    G_{a b} = \left(\frac{m_{\rm P}}{m}\right)_a \left(\frac{m_{\rm A}}{m}\right)_b G_{\rm N} \,,
\end{equation}
where $G_{\rm N}$ is the Newtonian gravitational constant, as measured in a Cavendish-type experiment, and $m_{\rm P}$ and $m_{\rm A}$ denote the passive and active gravitational mass respectively. For semi-conservative metric theories of gravity that have a conservation of momentum one has only a single gravitational mass $m_{\rm G} \equiv m_{\rm P} = m_{\rm A}$ \citep{Will_book}. For the remainder of the paper we assume that momentum is conserved in the gravitational interaction, and therefore $G_{ab} = G_{ba}$.\footnote{\cite{Shao_2016} investigates the possibility of constraining a difference in active and passive gravitational mass with the pulsar system under consideration in this paper.} More generally, we use the definition $G_{ab} = G_{\rm N} (1 + \Delta_{ab})$ where we denote $\Delta_{ab} = \Delta_{ba}$ as the relative GWEP parameter between two bodies $a$ and $b$.


If one observes an isolated two-body system without prior knowledge of the masses, then any violation of the GWEP at Newtonian order would be indistinguishable from a re-scaling of the masses due to the symmetry of the equations of motion.
This symmetry is broken in presence of a third body. One can then compare the rate at which two self-gravitating objects fall in the field of a third one. This forms the base for a class of GWEP tests that includes Lunar laser ranging (LLR), tests with binary pulsars falling in the gravitational field of our Galaxy, and the test to be discussed in this paper. 



In the LLR test, one considers the Earth-Moon system falling in the gravitational field of the Sun. If GWEP is violated, then the Earth, which has a larger fractional gravitational binding energy than the Moon, falls in the Sun's gravitational field with a slightly different acceleration than the Moon. This causes a polarisation of the Earth-Moon orbit in the direction of the Sun \citep{Nordtvedt_1968}. This so called Nortdvedt effect is the gravitational equivalent of the Stark effect, where a strong electric field polarises neutral atoms. It manifests itself as an added small orbital eccentricity vector that precesses in the sky with a period of 1 year, trailing the Sun. The relative Earth-Moon distance can be measured with an accuracy of about $10\,$cm thanks to the reflectors laid on the Moon by a variety of American and Soviet lunar missions. No Nodtvedt effect has been measured, as predicted by GR, effectively constraining
\begin{equation}
\Delta_{\rm E\odot} - \Delta_{\rm M\odot} \simeq 
    \left(\frac{m_{\rm G}}{m}\right)_{\rm E} - \left(\frac{m_{\rm G}}{m}\right)_{\rm M}
    = (-3.0 \pm 5.0) \times 10^{-14}
\end{equation}
\citep{Hofmann_2018}, which is only about a factor of 10 weaker than the MICROSCOPE limit for WEP, therefore confirming to a high degree that gravitational binding energy falls the same way in an external gravitational field as any other form of energy.

In this test, all the involved bodies are weakly self-gravitating, however, this is especially true for the two `proof masses', the Earth and the Moon: for the Earth $\varepsilon_\mathrm{grav, E} \equiv E_{\rm grav, E} / m_{\rm E}c^2 = -4.6 \times 10^{-10}$ (here $E_{\rm grav, E}$ is the Newtonian gravitational binding energy of the Earth), for the Moon $\varepsilon_\mathrm{grav, M} = -0.2 \times 10^{-10}$. This means that the LLR experiment only tests the weak-field limit of GWEP. In this limit Eq.~(\ref{eq:Gab}) implies $\Delta_{ab} \simeq \Delta_a + \Delta_b$, where $\Delta_a \equiv \left(m_{\rm G}/m\right)_a - 1$; furthermore, the gravitational binding energy of the bodies relative to their mass is so small that it can only have a very small effect on $\Delta_a$. Within the parametrised post-Newtonian (PPN) formalism for metric theories of gravity 
\begin{equation} \label{eq:Delta_a}
   \Delta_a = \eta \, \varepsilon_{{\rm grav},a} \,,
\end{equation}
where $\eta$ is the so called {\em Nordtvedt parameter}, a combination of several PPN parameters \citep[see][for details]{Will_book}.  The current limit on the Nordtvedt parameter from LLR is $(-0.2 \pm 1.1) \times 10^{-4}$.

A violation of GWEP not only affects the dynamics of the Earth-Moon system, but all self-gravitating masses in the Solar System are affected according to Eq.~(\ref{eq:Delta_a}). A consequence of this is a shift of the Solar System barycentre (SSB) when modelling planetary ephemerides. Based on data from the MESSENGER mission, \cite{Genova_2018} have derived $\eta = (-6.6 \pm 7.2) \times 10^{-5}$.

Equation \ref{eq:Delta_a} applies to the weak-field limit, that is, the Nordtvedt parameter parametrises GWEP violation to leading order in  $\varepsilon_{\mathrm{grav},a} \ll 1$. This first order approach is no longer applicable in the strong-field regime of neutron stars. Thus, in the remainder of this article we consider GWEP violations in terms of limits directly on $\Delta_{ab}$ and not on $\eta$.



In the Damour-Sch\"{a}fer test \citep{Damour_Schaefer_1991}, one verifies whether the two components of a pulsar - white dwarf system (the first with a very high degree of self-gravity, which allows the detection of strong-field SEP violation) fall with the same acceleration in the field of the Galaxy, which acts as the third body. A violation of the UFF would again cause a polarisation of the orbit of the binary pulsar. At the time of that paper (1991), the timing precision and timing baselines of binary pulsars were relatively small, so the authors proposed a statistical approach to search for this polarising effect in the orbital eccentricities of the known pulsar - white dwarf systems. Following that method, several analyses of the orbital eccentricities have constrained $\Delta$ for neutron stars:\footnote{More precisely, the constraint is on $\Delta_{\rm pulsar,Galaxy} - \Delta_{\rm companion,Galaxy}$.} \cite{Stairs_2005} derive $|\Delta| < 5.6\, \times \, 10^{-3}$, and \cite{gonzalez_high-precision_2011} derive $|\Delta| < 4.6\, \times \,10^{-3}$ (both being 95 \% confidence limits). However, the latter limit is derived with the inclusion of a binary pulsar, PSR~J1711$-$4322 that does not fulfil all the necessary criteria for the Damour-Sch\"{a}fer test \citep{Wex_2014}.

This method has several shortcomings, which are listed and discussed in detail by \citet{freire_tests_2012}; the two most important ones are a) the fact that it cannot detect GWEP violation, only produce statistical upper limits for it and b) generally, neutron stars with different masses have different values of $\Delta$, this limits the meaningfulness of a general $\Delta$ for neutron stars \citep[cf.\ footnote 25 in][]{Damour_2009}. 

Apart from the statistical test based on small eccentricities, \cite{Damour_Schaefer_1991} have also proposed a test based on a direct measurement of the variation of the orbital eccentricity vector for individual systems, $\dot{\vec{e}}$, (no matter whether eccentric or not) that results from the polarisation of the binary orbit by the Nordtvedt effect. As discussed by \cite{freire_tests_2012}, this test not only avoids all the shortcomings of the statistical test, but its precision just keeps improving with the precision of the measurement of $\dot{\vec{e}}$, which improves with time and with better timing instruments. Indeed, they estimated that this test should, for the best timed binaries, yield slightly better $\Delta$ values than the statistical test by the mid 2010's. More recently \cite{Zhu_1713} confirmed this by using the $\dot{e}$ constraint for the wide orbit of PSR~J1713+0747 to derive $| \Delta | <  2 \times 10^{-3}$ (95 \% C. L). Without further assumptions, this limit is strictly speaking only for neutron stars around $1.3 \, M_{\odot}$, which is the mass of PSR~J1713+0747. Recently, this limit has been used to test the UFF of a neutron star towards dark matter \citep{Shao_DM}.



Although this test can detect any hypothetical large strong-field deviations of the gravitational properties of neutron stars, the limits on $|\Delta|$ are not very constraining because of the weak gravitational field of the Galaxy, which has accelerations of the order of $2\, \times \, 10^{-10} \, \rm m \, s^{-2}$ in the solar vicinity. In the case of the LLR test, the polarising gravitational field (that of the Sun) is much stronger ($6\, \times \, 10^{-3}\, \rm m \, s^{-2}$), however, in that case the Earth and Moon have very small gravitational self energies.
 
For this reason, \cite{freire_tests_2012} suggested that the (then) rumoured pulsar in a triple system would combine the best features of both tests. In this experiment, we look for the Nordtvedt effect in an inner binary system consisting of a pulsar and a white dwarf; this system is orbited by a third hierarchical component significantly farther away. As in previous binary pulsar experiments, the pulsar provides the precise tracking and an object with very strong gravitational self energy; the white dwarf provides a test mass with a much smaller gravitational self energy, and finally the third outer component in that system provides a strong (potentially) polarising gravitational field ($1.7 \times 10^{-3}\,{\rm m\,s^{-2}}$), as the Sun does for the LLR experiment. The outer component would ideally be a neutron star, as this would provide a qualitatively different test, however, any type of star would already yield a much stronger polarising force than the Galactic gravitational field and therefore either a detection of GWEP violation, or much improved limits on it.



PSR~J0337+1715 was discovered in data from the GBT drift-scan survey  \citep{GBT_drift_1,GBT_drift_2}. This is a 2.7-ms pulsar in a 1.6-day orbit with a $\sim \, 0.2 \, M_{\odot}$ Helium white dwarf star, this is what we refer to from now on as the inner binary. The outer $\sim0.4\Msol$ white dwarf orbits the inner binary in about 327 days in a low-eccentricity ($e = 0.035$) orbit. This is the first, and so far the only pulsar confirmed to be in a triple stellar system \citep{ransom_millisecond_2014}. The two orbits (inner and outer) are nearly coplanar, this and the small observed eccentricities provide important clues for the evolution of the system, which was described in detail by \cite{Tauris_Triple}.

The pulsar has very good rotational stability, as usually millisecond pulsars (MSPs) do, and is relatively bright, which allows a very good measurement of the times of arrival of the pulses. This has allowed precise measurements of the varying orbital parameters, and also extremely precise mass measurements for the pulsar and the two white dwarf stars \citep{ransom_millisecond_2014}. More importantly, the GWEP test was eventually carried out for PSR~J0337+1715 by \cite{archibald_universality_2018}, yielding $|\Delta| \leq 2.6 \times 10^{-6}$ (95\% C.L.). This represents an improvement of three orders of magnitude over previous pulsar tests and confirmed the power of a MSP in a triple stellar system for testing the GWEP.

This measurement represents a tight constraint on alternative theories of gravity. \cite{archibald_universality_2018} derived constraints on one of the best studied alternatives to GR, the class of mono-scalar-tensor theories described by Damour \& Esposito-Far\`ese (1992, 1993\nocite{DEF_1992,DEF_1993}, henceforth DEF gravity). The constraints on the weak-field coupling parameter for these theories ($\alpha_0$) derived from PSR~J0337+1715 significantly improve upon all previous experiments for most of their $\beta_0$ space.


The UFF experiment with the PSR~J0337+1715 triple system and its results are clearly of great importance. It is, at present, the most powerful test of the GWEP, for either the strong or weak field limits. It is also extremely sensitive to strong-field deviations in the gravitational properties of neutron stars. 

This test is of special value because, according to a gravitational analogue of Schiff's conjecture, it is plausible that the validity of GWEP implies the SEP \citep{Will_book}; this in turn strongly suggests, according to the arguments mentioned above, that GR is {\em the} theory of gravity \citep[see also][]{will_confrontation_2014}.

For all these reasons, we find it is important to improve both the precision and reliability of the test. These are the primary objectives of this work. We use fully independent observational data, taken with a wholly different observing system (described in detail in Section~\ref{sec:Observation}), a completely independent numerical integration of the motion of the system and a different implementation of the determination of the masses and orbital parameters (described in Section~\ref{sec:numerical_model}) than those used by \cite{archibald_universality_2018}. One of the main differences is, however, that the uncertainties we report are purely statistical; we found no need to postulate the existence of additional systematic effects that can bias $\Delta$. Consequently, this parameter can be self-consistently processed like the others without requiring a special treatment.

The results of our analysis are presented in Section~\ref{sec:parameters}. Here we discuss not only the parameters we obtain, but also analyse the trends observed in the residuals after subtracting the best model for the system. In Section~\ref{sec:tests_theory}, we interpret the $\Delta$ constraint, as well as constraints on the post-Newtonian strong-field parameters, within the context of a wide framework of alternative theories of gravity, the Bergmann-Wagoner theories of gravity \citep{Will_book}. We also derive new constraints on a sub-class of those theories, the Damour-Esposito Far\`ese (DEF) theories \citep{DEF_1992,DEF_1993}, these new limits are derived in a conservative way that accounts for uncertainties in our knowledge of the equation of state (EOS) for neutron star matter. We finally summarise our findings in Section~\ref{sec:summary}.




\section{Observations and data reduction \label{sec:Observation}}


The pulsar J0337+1715 has been regularly observed since July 2013 every 2 or 3 days with the Nan\c cay radio telescope using its L-band receiver at a central frequency of 1484\,MHz.
The Nan\c cay radio telescope is a meridian Kraus design collector equivalent to a 94-meter dish able to conduct $\sim$1 hour observations on any given source within its declination range each day. The dual linear polarisation signals are sent to the Nan\c cay Ultimate Pulsar Processing Instrument (NUPPI, \citealt{Desvignes2011}), an instrument that is able to coherently dedisperse \citep{1975mcpr...14...55H} a total bandwidth of 512\,MHz. It consists of a ROACH1 board (designed by the CASPER group, University of California, Berkeley) providing 128 baseband data streams of 4MHz bandwidth each. The instrument software has many similarities with GUPPI (Green Bank Ultimate Pulsar Processing Instrument, \citealt{2008SPIE.7019E..1DD}) used at the Green Bank Telescope (GBT). A cluster of four nodes hosting eight GTX280/285 Graphics Processing Units (GPUs) is used to coherently dedisperse and fold the data in real-time.

The real-time folding process uses a pulsar period coming from a simple model with two non-interacting Keplerian orbits over short 15-second sub-integrations. A single standard timing parameter file in \textsc{tempo} format\footnote{\textsc{tempo} is a standard pulsar timing software, this can be found at \url{http://tempo.sourceforge.net}.} containing this pulsar timing model is used for all the observations. The full frequency and time resolution daily pulsar profiles are stored in \textsc{PSRFITS} archives \citep{psrchive}\footnote{\url{http://psrchive.sourceforge.net}.}. A $\sim$3\,Hz pulsed noise diode is fired for $\sim$10 seconds at the start of each observation to conduct a simple calibration accounting for gain and phase differences between the two polarisations, as implemented in the \textsc{singleaxis} polarisation calibration of \textsc{psrchive}.

\begin{figure}[htb]
    \centering
    \includegraphics[width=\columnwidth]{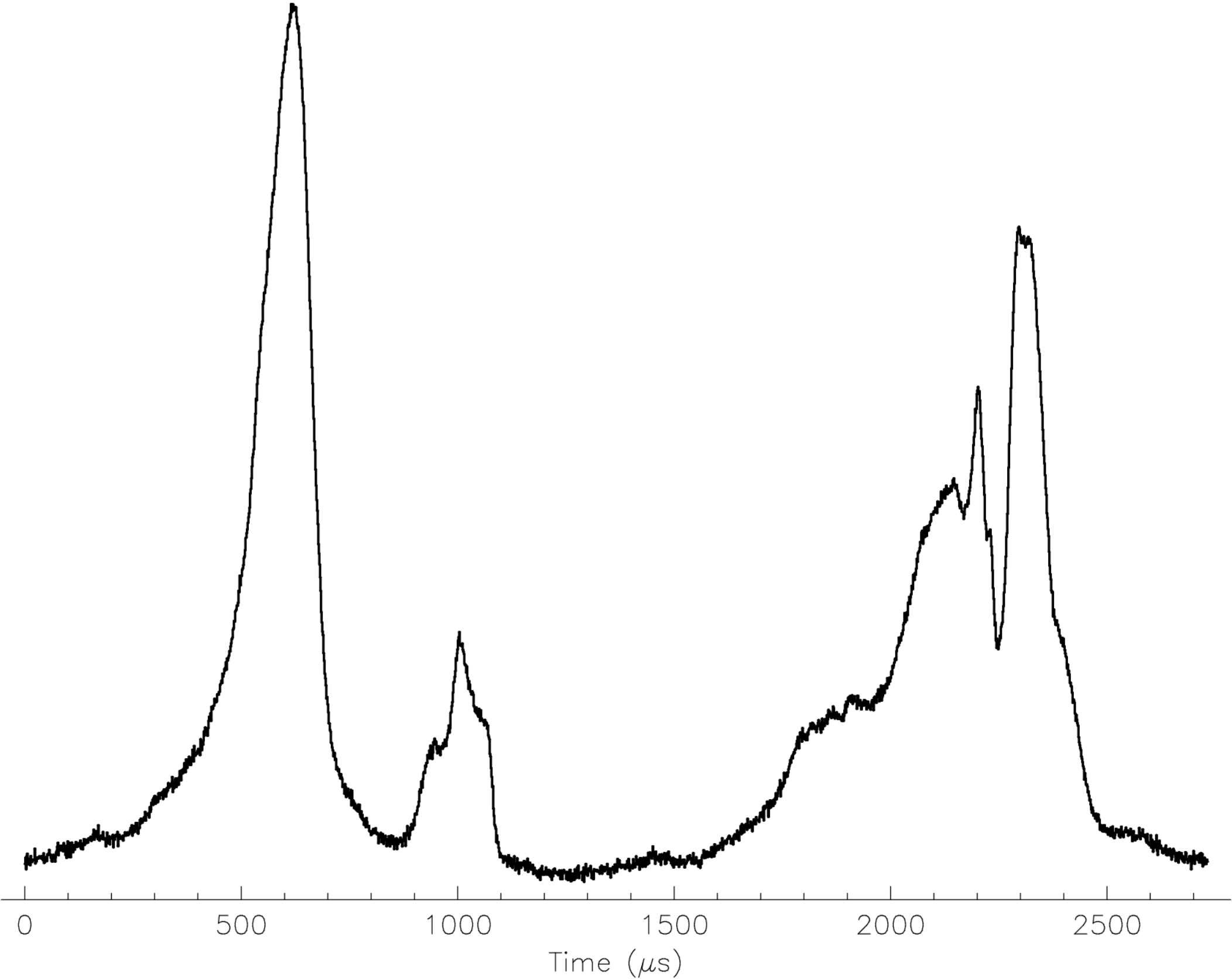}
    \includegraphics[width=\columnwidth]{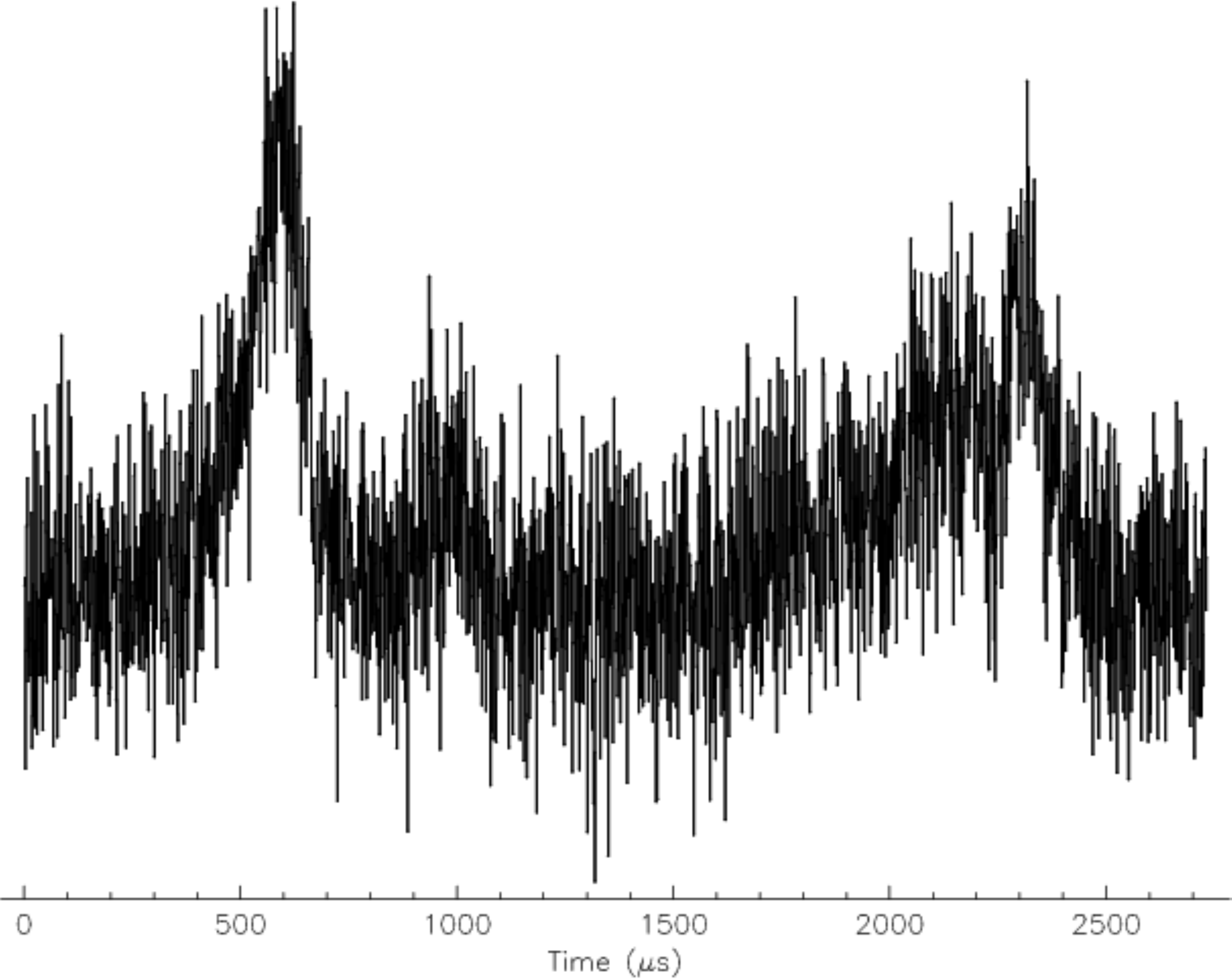}
    \caption{(top) Template pulse profile of PSR J0337+1715 used for timing. 450 hours of observations conducted with the Nan\c cay Radio Telescope were integrated over the frequency range 1230--1742\,MHz. (bottom) Pulse profile obtained on October 4, 2014. When profiles are ToA uncertainty ranked, this one is at the 10th percentile of the lowest uncertainties, typical of a good observation.}
    \label{fig:template}
\end{figure}

As the pulsar period model used to fold in real time is not strictly satisfactory, it is necessary to properly phase-shift all the archived individual profiles. A posteriori, the drifts within individual sub-integrations were statistically smaller than the mean ToA uncertainty and characterised by an rms of 0.78$\,\mu$s, thus validating the parameters of our sub-integrations. In an iterative way, measured times of arrival are used to derive a pulsar timing model which is used to improve the times of arrival and so on. Practically, the numerically derived pulsar timing model is used to provide `theoretical' barycentric arrival times for each of the 15-second sub-integrations. A code transforms those barycentric arrival times into a simple daily \textsc{tempo} parameter file with rotation rate described by only a frequency and its first three derivatives around an epoch corresponding to the middle of the observation (F0, F1, F2, F3 and PEPOCH). This polynomial predicts the barycentric rotational phases within 5\,ns at maximum. These daily parameter files are then used to re-align the pulse profiles within their corresponding archives. A profile `template', built with more than one thousand observations (see Fig. \ref{fig:template}, top), is used for determining the topocentric pulse times of arrival (ToAs) in the following way. After integrating profiles over 128 MHz and 20 minutes, the times of arrival are estimated using the \texttt{pat} function from \textsc{psrchive} within the Fourier domain with Markov chain Monte Carlo (FDM) method. The total bandwith of 512\,MHz was thus split in four sub-bands in order to be able to fit for variations in the dispersion measure (DM) representing the integrated electron column density along the line-of-sight during the numerical fit. The ToA uncertainties as reported by \verb,pat, are characterised by a mean of 2.15\,$\mu$s and a median of 1.89\,$\mu$s. A pulse profile typical of a good observation, characterised by an uncertainty of 1.15\,$\mu$s, is shown in Fig. \ref{fig:template} (bottom). The goodness of fit as reported by \texttt{pat} can give a sense of the differences between the template profile and the profiles used to derive ToAs. The goodness values are characterised by a mean of 1.05 (with a median of 1.04) and an r.m.s. of 0.12 with 99\% of the values between $\pm$3$\sigma$ (0.69 to 1.41). The rather low signal to noise ratio of the PSR~J0337+1715 profiles observed at Nan\c cay prevents the detection of subtle effects of incorrect polarisation calibration on the ToA determination.
In this work, we use a dataset (see footnote \ref{fn:zenodo}) of 9303 ToAs divided in four 128\,MHz bands observed between MJD 56492 and MJD 58761 (July 2013 and October 2019).
 

\section{\label{sec:numerical_model}NUmerical TIming MOdel: NUTIMO}

For the description of the orbits of binary pulsars, timing programmes such as the aforementioned \textsc{tempo} and also \textsc{tempo2} \citep{edwards_tempo2_2006, hobbs_tempo2_2006} rely on existing analytical models to calculate the times of arrival with nanosecond accuracy (such as, e.g. the DD and DDGR models, \citealt{damour_general_1986}). These models are built from precise analytical solutions of the equations of motion (for the examples above these were derived by \citealt{damour_general_1985}). However, no such solution is yet available for a triple system where the three masses are of comparable size and experience moderately strong mutual interactions. Therefore, and similarly to \cite{ransom_millisecond_2014} and \cite{archibald_universality_2018}, we perform a high precision numerical integration of the equations of motion, which we subsequently use to calculate the delays.

The equations of motion we use are accurate up to first post-Newtonian order (1PN), that is,\ include the first-order terms of an expansion of GR in the small parameter $\epsilon \sim \frac{Gm}{a c^2} \sim \frac{v^2}{c^2} \lesssim 5\times 10^{-7}$  where $m$, $v$ and $a$ are characteristic mass, velocity and length scales of the system and $c$ the speed of light. These corrections are absolutely necessary because they translate into a relative acceleration $\Delta a/ a \sim \epsilon$ which is of similar magnitude as a potential SEP violation (see above). In addition, 1PN corrections are responsible for effects that accumulate over time such as the well-known relativistic precession of periastron. On the other hand, second order corrections can be safely ignored since the same line of reasoning predicts that even a cumulative effect such as gravitational wave radiation cannot account for more than a nanosecond within the current span of our observations.

We use the 1PN generic strong-field framework of \cite{will_theory_1993} and \cite{damour_strong-field_1992} which parametrises almost the entire class of `fully conservative' Lagrangian-based  theories of gravity (without preferred location effects\footnote{Preferred location effects are already tightly constrained using pulsars by \cite{Shao_LPI}.}) based on a modified Einstein-Infeld-Hoffmann approach (see details in Appendix \ref{sec:eqofmotion}). In this framework, in the most generic case, one has three parameters $\Delta_{ab}$ at the Newtonian and 12 strong-field parameters at the post-Newtonian level. All these parameters depend on the structure of the individual bodies. The 12 post-Newtonian parameters generalise the parametrised post-Newtonian (PPN) $\beta_{\rm PPN}$ and $\gamma_{\rm PPN}$ \citep{will_theory_1993} to the regime of strongly self gravitating masses. 

In the PSR~J0337+1715 system we have only one strongly self-gravitating body with $\varepsilon_{\rm grav} \sim 0.1$, the pulsar, while the two white dwarfs are weak-field objects with $\varepsilon_{\rm grav} \lesssim 10^{-4}$. That generally leads to a significant reduction of the number of strong-field parameters relevant for the orbital dynamics of the PSR~J0337+1715 system. In fact, on the Newtonian level there is only one $\Delta_{ab}$, which we simply denote by $\Delta$. Among the post-Newtonian terms, as we discuss in detail within a theory based framework in Section~\ref{sec:tests_theory}, there remain three strong field parameters which are a priori unconstrained by Solar System experiments and limits on $\Delta$ already imposed by the `Newtonian' dynamics: $\bar{\beta}_{\rm p}$, $\bar{\beta}_{\rm pp}$, $\bar{\beta}^{\rm p}$. Since the limits we find for these parameters in Section~\ref{sec:parameters} are many orders of magnitude weaker than limits inferred indirectly from binary pulsar experiments, at least within our theory based framework, we primarily consider a model where the 1PN strong-field parameters are set to their GR values that is, zero. This practically corresponds to using priors arising from a combination of Solar System and binary pulsar limits at the post-Newtonian level  when estimating $\Delta$.

Our specially developed software, \textsc{nutimo}\footnote{Source, data, and results are available at: \url{doi.org/10.5281/zenodo.3778978} \label{fn:zenodo}}, solves numerically the 3-body equations of motion at 1PN (see appendix \ref{eqtr:lagrangian1pn} and particularly equation \eqref{eqtr:motion}) before computing propagation and relativistic delays. All the geometrical delays are taken into account up to first order in $L/d$ where $d$ is the distance to the system and $L \ll d$ is any other length scale of the problem. In other words, the code computes the so-called R\oe mer, Kopeikin and Shklovskii delays \citep{shk70,kopeikin96}, and adds an extra second order correction, that is, $L^2/d^2$, for the latter (the only second order correction that may become important with time, see e.g. \citealt{voisin_simulation_2017}). We note that Kopeikin and Shklovskii delays were not included in previous works \citep{archibald_universality_2018, ransom_millisecond_2014}. The former allows us to measure the longitude of ascending node of the outer orbit and might be important because it accounts for systematic effects at the Earth orbital frequency which is close to the outer orbital frequency. We do not expect the latter to significantly affect the results of the fit but it allows us to derive the intrinsic pulsar spin parameters which would otherwise absorb this effect (see below). We caution that the intrinsic spin parameters we report in Table \ref{tab:fitresults} are still biased by the effect of Galactic acceleration which amounts to approximately 25\% of the Shklovskii correction (according to the Galactic model of \cite{McMillan_2017}). Relativistic delays include time dilation between the frame of the pulsar and the frame of the observer, namely the so-called Einstein delay, as well as the deformation of space-time by the pulsar companions on the light travel path, the so-called Shapiro delay, and the aberration of the direction of the radio beam. All are calculated at 1PN order.

The delays due to interstellar medium propagation described by the DM as well as the Solar System counterparts of the previously mentioned delays are calculated by the commonly used software \textsc{tempo2} \citep{edwards_tempo2_2006,hobbs_tempo2_2006} which \textsc{nutimo} integrates as an external library. A thorough description of the timing model can be found in Chapter 5 of \cite{voisin_simulation_2017}. 


\subsection{Parametrisation of the problem \label{sec:param}}

In total, the model must include at least 27 parameters (respectively 30 in the so-called secondary model when the 3 1PN strong-field parameters are included).
One of them is a ToA uncertainty scale factor (called EFAC in the pulsar timing literature) which quantifies our ignorance of unmodelled systematic effects. The other 26 parameters (resp. 29) can be grouped into four categories:
\begin{itemize}
    \item pulsar rotation: pulsar spin frequency and its derivative;
    \item orbital dynamics: six parameters for the inner-binary orbit, six parameters for the outer-WD orbit, three masses, one SEP violation parameter (resp. one SEP violation parameter and three 1PN strong-field parameters);
    \item astrometry: three position and three proper motion parameters;
    \item radio propagation: DM and DM derivative. 
\end{itemize}
Each category is essentially uncorrelated with the others (see Figure \ref{fig:corner}). The first two are specific to the triple-system problem and we shall discuss them in some details. On the other hand, the astrometric parameters, position and proper motion, and DM are treated using a standard approach and we refer the interested reader to \citet{edwards_tempo2_2006}, for example. 

The intrinsic pulsar parameters are its spin frequency $f$ and spin-frequency derivative $f'$ taken at the reference epoch $T_\mathrm{ref}$. These parameters need to be re-scaled to avoid non-linear correlations with astrometric parameters due to the Shklovskii delay (see Chapter 5 of \cite{voisin_simulation_2017} for details about the delays). This is a common practice in pulsar timing. In addition, we re-scaled the spin frequency to include the linear effect (that is, proportional to time) of the Einstein delay, which is approximated by the second term of Eqn. \eqref{eq:fbar} below. In usual pulsar-timing models, the term of the Einstein delay responsible for a linear increase of the delay with time can be calculated exactly and removed from the timing model since its effect is strictly impossible to separate from a re-scaling of $f$. However, because here we calculate numerically the delay, we can only estimate the linear drift using the initial parameters. As a result of these re-scaling, the effective fit parameters  $\bar{f}$ and $\bar{f}'$ are connected to the intrinsic parameters $f$ and $f'$ by 
\begin{eqnarray}
\label{eq:fbar}
 \bar{f} & = & f \left[ 1 - \frac{v^2 + U_{\rm i} + U_{\rm o}}{2c^2} - \mu_\perp^2 d \Delta T \left( 1 - \frac{3}{2}\Delta T \mu_{\rm d} \right) \right] / c \, ,\\
 \bar{f'} & = & f' - f d \mu_\perp^2 (1 - 3 \Delta T\mu_{\rm d}) /c \, ,
\end{eqnarray}

where
\begin{eqnarray}
v & = & 2\pi \left( \frac{a_{\rm b}}{P_{\rm O}} + \frac{a_{\rm p}}{P_{\rm I}} \right) \, , \\
U_{\rm i} & = & \frac{G m_{\rm i}}{a_{\rm p} (1 + m_{\rm p}/m_{\rm i})} \, , \\
U_{\rm o} & = & \frac{G m_{\rm o}}{a_{\rm b} (1 + (m_{\rm p}+m_{\rm i})/m_{\rm o})} \, , \\
\Delta T & = & T_{\mathrm{ref}} - T_{\mathrm{pos}} \, ,
\end{eqnarray}
where the symbols correspond to those defined in Table \ref{tab:fitresults}. The use of the two re-scaled parameters above instead of the intrinsic ones has proven to be very effective in speeding up convergence in our MCMC.

\begin{figure}[htb]
    \centering
    \includegraphics[width=\columnwidth]{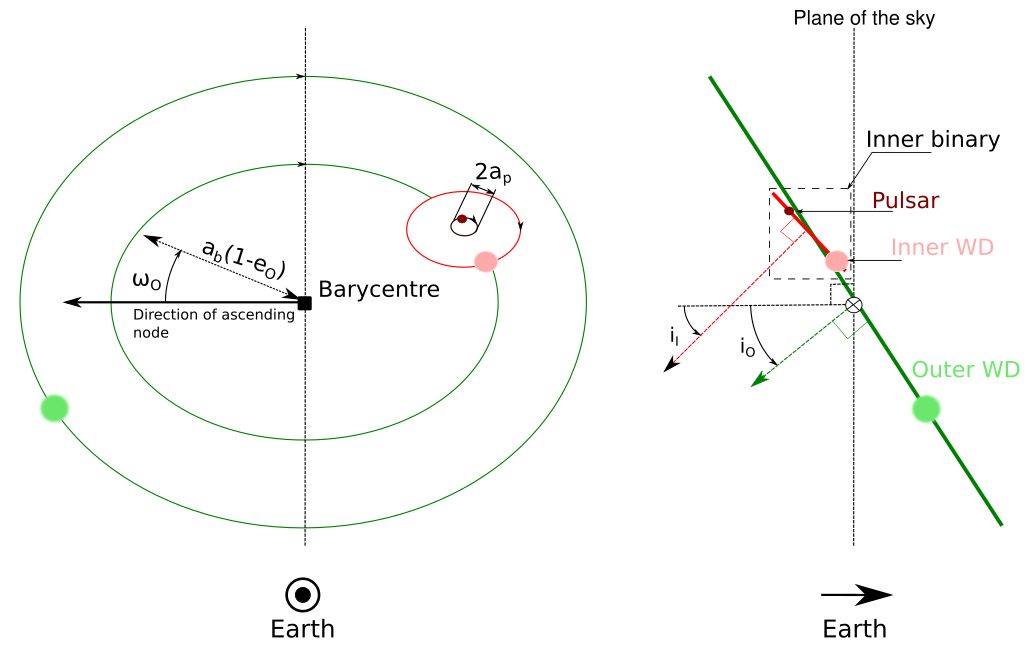}
    \caption{Sketch of the orbits of PSR J0337+1715 (not to scale). Note that the orbits are nearly circular and that the ellipsoidal shape of the orbits on the left-hand sketch arises purely from projection. The osculating orbits of the system can be parametrised by the Keplerian orbital elements of the pulsar within the inner binary and of the inner binary within the outer binary (see text). In particular, $a_{\rm p}$ is the semi-major axis of the pulsar orbit within the inner binary, $a_{\rm b}$ is the semi-major axis of the inner binary within the outer binary and $e_{\rm O}$ its eccentricity. Are also shown the longitude of periastron the outer binary as well as the inner and outer inclinations with respect to the plane of the sky, $i_{\rm I}$ and $i_{\rm O}$ respectively. Note that for simplicity the sketch neglects the difference between the directions of the inner and outer lines of ascending nodes, $\delta\Omega$, as it is in practice very small.}
    \label{fig:orbits}
\end{figure}
The orbital parameters for a triple system are at most $3\times 6 + 3 = 21$ that is, three position coordinates and three velocity coordinates per body plus the three masses. However we consider the system in the frame of its centre of mass which results in applying two vector relations to the initial velocities and positions such that the centre of mass is at rest at coordinates $(0,0,0)$. These relations suppress six degrees of freedom, and we end up with fifteen independent orbital parameters. Note that, eventually, the six degrees of freedom corresponding to the centre of mass appear as the six astrometric parameters. 

The orbital parametrisation uses the fact that the system, being hierarchical, can approximately be described by an inner Keplerian binary containing the pulsar and the inner white dwarf (WD) itself forming an outer binary with the outer WD (see Figure \ref{fig:orbits}). Thus, the usual Keplerian orbital elements can be used to describe the osculating Keplerian orbits to the actual trajectory of the pulsar and of the inner binary.  For eccentricity, we use the Laplace-Lagrange parametrisation relevant for small eccentricities \citep{lange_precision_2001} which replaces $e, \omega, t_p$, respectively eccentricity, longitude of periastron and time of periastron passage,  by $e\cos\omega, e\sin\omega$, and $t_\mathrm{asc}$. It is important to note that we define the transformation $t_\mathrm{asc} = t_p - \omega/2\pi P$, with $P$ an orbital period.

Similarly, we fitted for $a_{\rm b}\sin i_{\rm O}, a_{\rm b} \cos i_{\rm O}$ for the outer orbit, where $a_{\rm b}$ is the semi-major axis of the inner binary's orbit around the centre of mass of the system and $i_{\rm O}$ its inclination relative to the plane of the sky. 
For the inner binary we find it better to fit for $a_{\rm p} \sin i_{\rm I}$ and $\delta i = i_{\rm I} -i_{\rm O}$ instead of $a_{\rm p} \sin i_{\rm I}$ and $a_{\rm p} \cos i_{\rm I}$ as this cancels several non-linear correlations in the fit. This is helped by the fact the two orbits are very nearly coplanar.
In the same way, we fitted for the longitude on the plane of the sky of the outer orbit, $\Omega_{\rm O}$, as well as for the offset between inner and outer orbits, $\delta \Omega = \Omega_{\rm I} -\Omega_{\rm O}$. We note that only the latter was reported in \citet{archibald_universality_2018} while the former was considered impossible to constrain with the available data. Interestingly, we were able to constrain $\Omega_{\rm O}$ in this work, perhaps thanks to our inclusion of Kopeikin's delays.

The inner binary mass $m_{\rm b} = m_{\rm i} + m_{\rm p}$ as well as the outer WD mass are derived using Kepler's third law in the inner and outer binary respectively. The pulsar mass and the inner WD mass are derived from $m_{\rm b}$ using the mass ratio $m_{\rm i}/m_{\rm p}$ which is also part of the fit. In addition, we use the post-Keplerian orbital elements of \cite{damour_general_1985} which incorporate the 1PN corrections for relativistic binaries. Using \cite{damour_general_1985}, one maps the orbital elements to the position and velocity of the pulsar relative to the inner-binary centre-of-mass $(\vec{r}_{\rm p/b}, \vv_{\rm p/b})$ and to the inner-binary centre-of-mass position and velocity $(\vec{r}_{\rm b}, \vv_{\rm b})$. One can then find the position and velocity of the pulsar relative to the centre of mass of the system, $\vec{r}_{\rm p} = \vec{r}_{\rm p/b} + \vec{r}_{\rm b}$ and $\vec{v}_{\rm i} = \vec{v}_{\rm i/b} + \vec{v}_{\rm b}$.\footnote{Although addition of velocities only applies to Newtonian mechanics, we are here only interested in transforming the orbital elements into initial conditions for the numerical integrator. Such transformation is somewhat arbitrary, and we choose to add these velocities for simplicity.} The position and velocity of the inner WD, $(\vec{r}_{\rm i}, \vv_{\rm i})$, can then be derived by solving the equations of conservation of momentum and centre-of-mass position, \eqref{eqtr:momentum} and \eqref{eqtr:cdm1pn} respectively, with $\vec{P} = (\mathscr{H}/c^2) \vec{v}_{\rm b}$ and $\vec{X} = \vec{r}_{\rm b}$. Finally, one solves $\vec{P} = 0$ and $\vec{X} = 0$ for the outer WD position and velocity, $(\vec{r}_{\rm o}, \vv_{\rm o})$.


\subsection{Model accuracy}

The main signature of a SEP violation in our pulsar-timing experiment would be a residual signal at the frequency $2f_{\rm i} - f_{\rm o}$ \citep{archibald_universality_2018}, where $f_{\rm i,o}$ are the inner and outer orbital frequency respectively (see also Figure \ref{fig:periodogram}). According to linear theory, the effect of a violation of the SEP at Newtonian order primarily results in a sinusoidal variation of the separation of the inner binary with frequency $f_{\rm i} - f_{\rm o}$ \citep{Nordtvedt_1968}. However, here we measure the distance projected along the line of sight between the observer and the pulsar and this distance is modulated by the orbit of the inner binary with the outer WD. It follows that the main net effect is a modulation at $2f_i - f_o$  as originally pointed out by \cite{archibald_universality_2018}. Using the Newtonian-order linear theory of \citet{Nordtvedt_1968}, one can show that $|\Delta| \sim 10^{-6}$ translates into a signal amplitude of $\sim 0.1\,$s via the R\oe mer delay. However, the magnitude that can effectively be detected in the fit residuals appears to be reduced to only $\sim50\,$ns \citep{archibald_universality_2018} due to the effect of the many strong correlations of the $\Delta$ parameter with the other orbital parameters (see Figure \ref{fig:corner}).
Therefore, we aim in this work to achieve nanosecond accuracy within our model. This level of accuracy is compatible with the level aimed at by \textsc{tempo2} \citep{edwards_tempo2_2006, hobbs_tempo2_2006}. 

There are essentially three types of inaccuracies that can affect the output of our model:numerical round-off errors, post-Newtonian truncation, interpolation precision. 
The first one, numerical round-off errors, is expected to grow with the number of floating-point operations performed to obtain the result. The main source of operations is the numerical integration of the equations of motion \eqref{eqtr:motion}. The integration is performed using the Bulirsch-Stoer scheme \citep{stoer_introduction_2011} implemented in the Odeint module \citep{ahnert_odeint_2011} of the Boost library \footnote{Boost library version 1.55.0 \url{www.boost.org}}. In addition our numerical model relies on 80 bit floating point numbers (long double in C) throughout. To assess the effect of numerical round-off errors, we use the model to generate fake times of arrival from parameters that fit the data from PSR J0337+1715. The fake times of arrival are the theoretical times of arrival that are closest to the actual measurements, and therefore only differ from those by a few microseconds at most. We feed this mock data set back into our model such that the residuals should be exactly zero in absence of numerical round-off errors. In practice we observe round-off errors at the level of $10^{-3}$ns showing that numerical round-off errors are not an issue given our objective of a 1ns accuracy. Note that this procedure does not assess any systematic inaccuracy due to the modelling or the numerical scheme themselves (see below) but it does account consistently for the entire chain of operations, including not only the numerical integration but also the Solar-system calculations done by \textsc{tempo2} and the pulsar system delays. It is also conservative since the chain of operations is performed twice: once to create the fakes and once to analyse them. 

The main source of systematic inaccuracy due to numerical approximations lies in the precision of the interpolations of the timing delays that are calculated in intermediate steps. We use a cubic-spline interpolation algorithm \citep{press_numerical_1996} for all our interpolations. There are two parameters than can be tuned: the number of interpolation points and the width of `margins' at the beginning and the end of the interpolated range in order to avoid boundary effects. The latter need only be a few points in principle, however the former has a direct and opposite impact on accuracy and performance and therefore requires a trade-off. To determine the level of interpolation accuracy required we need to estimate what is the effect of a 1ns signal on the $\chi^2$ value in order to make sure that this value is computed with the necessary accuracy. Let us assume that the difference between the data and the model is $\Delta t_i +\delta_i$ where the second term explicitly represent the contribution of a putative $\sim 1$ns signal, then the $\chi^2$ can be expanded as follow,
\begin{equation}
\label{eq:chi2delta}
    \chi^2 = \sum_{i=1}^N \frac{\Delta t_i^2}{\sigma_i^2} + \sum_{i=1}^N \frac{\Delta t_i \delta_i}{\sigma_i^2} + \sum_{i=1}^N \frac{\delta_i^2}{\sigma_i^2},
\end{equation}
where $N$ is the number of data points. 
Now if $N\gg 1$ and the number of fit parameters is $\ll N$, then for the best fit parameters $\Delta t_i \sim \sigma_i$ where $\{\sigma_i\}_{i\in [1,N]}$ are the uncertainties which we take to be approximately equal to $\sigma = 2\,\mathrm{\mu s}$. In the present case $N\sim 10,000$ for only 27 parameters. Thus, the first term in \eqref{eq:chi2delta} approximates $\chi^2 \sim N$. The second term in equation \eqref{eq:chi2delta} can be as large as $N \delta/\sigma$ assuming that every term contributes positively. However it is also possible that the sum averages to zero if it alternates. The third term is of order $\sim N \delta^2/\sigma^2$. Taking $\delta_i \sim \delta = 1\,$ns we see that the second and third term of equation \eqref{eq:chi2delta} are respectively $\lesssim 5\times 10^{-4}\chi^2$ and $\sim 2.5\times 10^{-7}\chi^2$. We retain the last estimate as a conservative level of accuracy to achieve. To do so we increased exponentially the number of interpolation points until the relative variation of the $\chi^2$ between two increments is smaller than $2.5\times 10^{-7}$.

\begin{figure}[htb]
    \centering
    \includegraphics[width=\columnwidth]{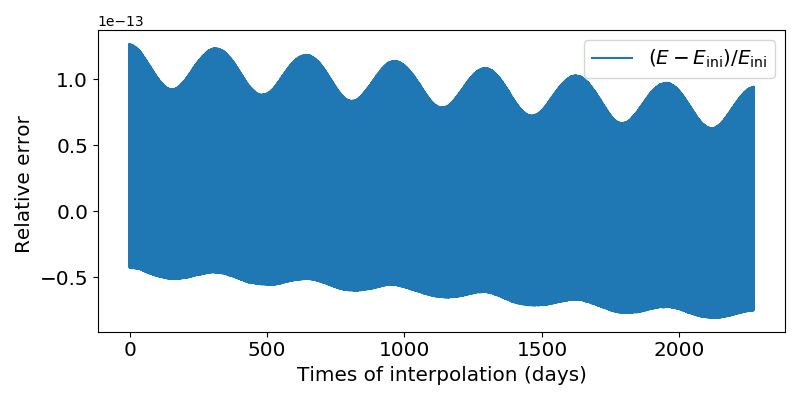}
    \caption{Relative error in energy conservation during the numerical integration of the equations of motion over the $2268$ day span of our data. The envelope of the signal  oscillates at the outer binary frequency while a zoom would show a fast oscillation at the inner binary orbital frequency.}
    \label{fig:nrjcons}
\end{figure}

\begin{figure}[htb]
    \centering
    \includegraphics[width=\columnwidth]{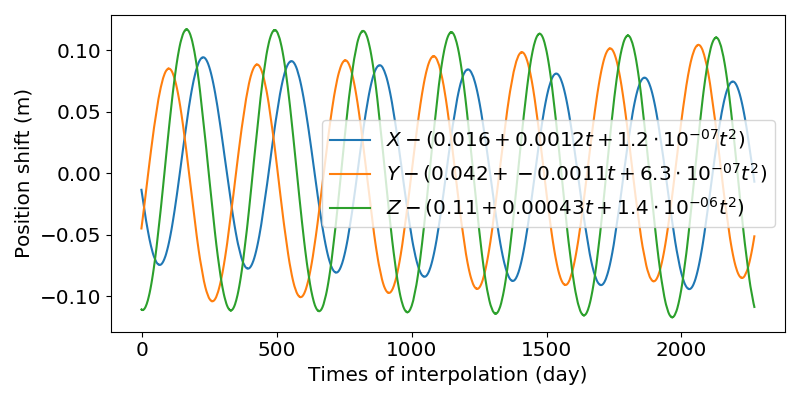}
    \caption{Three components of the centre-of-mass position variation during the integration of the equations of motion over the $2268$ day span of our data. A quadratic component has been fitted out (see text and legend) which leaves only the oscillatory components. The main visible oscillation is at the frequency of the outer binary while a zoom would show a fast oscillation at the inner binary orbital frequency.}
    \label{fig:cofcons}
\end{figure}
The last source of inaccuracy, post-Newtonian truncation, is intrinsic to the theoretical framework used. Indeed, although the equations of motion and the various conserved quantities of Section \ref{sec:eqofmotion} all consistently derive from the same Lagrangian and can therefore be exactly verified in principle, the method of calculus by successive approximation does not in practice achieve that result. Indeed, since the Lagrangian itself is an approximation to order $v^2/c^2 \sim 5\times 10^{-7}$ of a more general theory, there is no physical justification for conserving in the subsequent derivation any term of higher order. It follows that the equations of motion and the corresponding conserved quantities are only accurate to first order (1PN) and that systematic `residuals' of order $v^4/c^4 \sim 2\times 10^{-13}$ (2PN) are present in the equations themselves. As a result, we see on Figure \ref{fig:nrjcons} that the energy of the system is conserved up to systematic oscillations at the orbital frequencies accompanied with a linear drift at a level consistent with the neglected 2PN terms, and numerical noise does not play any significant role. More interesting regarding the timing accuracy is to look at the conservation of the centre-of-mass position. Indeed, a shift in this position immediately transforms into a geometric delay. We find that, due to the fact that the neglected 2PN terms in the equations of motion do not necessarily generate residuals which average to zero, the two successive integrations $\ddot x \rightarrow \dot x \rightarrow x$ leading to the centre-of-mass motion create a quadratic drift that increases over time. Through the time span of our observations this results into a drift of less than 10\,m, namely about 3\,ns in terms of geometric delay. Such a quadratic drift can undoubtedly be entirely absorbed in the spin frequency and spin-frequency derivative as well as by the astrometric parameters when fitting the data. For example, the linear drift reported on Figure \ref{fig:cofcons} would bias the spin frequency by $\sim 10^{-14}\,$Hz, much less than the nevertheless very tight uncertainty on this parameter. Therefore we conclude that the quadratic drift can only result in a negligible bias of a few parameters which is why we subtract this component with a linear-least-square fit on Figure \ref{fig:cofcons}. The residuals show that the systematic oscillatory 2PN motion of the centre of mass does not exceed $\sim 0.1\,$m, or $0.3\,$ns in terms of geometric delay, which is well within our tolerance.

In practice, the largest systematic errors may come from unmodelled effects. In particular, gravitational wave damping in the inner binary should account for a few nanoseconds after 5 years. Another effect that might become important for high-precision timing over time is the effect of the gravitational quadrupole moment of the inner white dwarf. Indeed this star should be slightly deformed by the tidal field of the neutron star and by its spin which would lead to a slight correction to the orbital precession rate.  


\section{Bayesian analysis results}
\label{sec:parameters}

\begin{table}
\caption{Gaussian priors adopted in MCMC posterior inference.}
    \centering
    \begin{tabular}{cllc}
         Parameter & Mean & Std dev  & Source\\
         \hline
         $\alpha$ & $3^\mathrm{h}37^\mathrm{m}43^\mathrm{s}.8270$ & $0.37\,$mas$^\dagger$ & 1\\
         $\delta$ & $17^\circ15'14''.8178$ & $0.38$\,mas$^\dagger$ & 1\\
         $d$ & $1.3$\,kpc & $160$\,pc$^\dagger$ & 3\\
         $\mu_{\rm \alpha}$ & $4.8\rm \, mas\, yr^{-1}$ & $1 \rm \,mas\, {yr^{-1}}^\dagger$ & 1 \\
         $\mu_{\rm \delta}$ & $-4.4\rm \, mas\, yr^{-1}$ & $0.8\rm \,mas\, {yr^{-1}}^\dagger$ & 1\\
         $V_\parallel = \mu_{\rm d} d$ & $29.7\rm \, km \, s^{-1}$ & $0.9 \rm \,km \, {s^{-1}}^*$ & 2\\
         \hline
         \multicolumn{4}{l}{$^\dagger$: $2\times$ the uncertainty reported in the source.} \\
         \multicolumn{4}{l}{$^*$: $1\times$ the uncertainty reported in the source.} \\
    \end{tabular}
    \tablebib{(1) Gaia DR2 \citep{lindegren_gaia_2018}; (2) \citet{kaplan_spectroscopy_2014}; (3) \citet{ransom_millisecond_2014}.}
    \label{tab:priors}
\end{table}

\begin{table*}
\caption{\label{tab:fitresults} Mean values of the MCMC fit with their $68\%$ confidence interval. }
\centering
		\renewcommand{\arraystretch}{1.15}
\begin{footnotesize}
\begin{tabular}{lcl}
Parameter & Symbol & 
\\
\hline
\multicolumn{3}{c}{Fixed values} \\
\hline
                Reference epoch ($\mathrm{MJD}$) & $T_{\mathrm{ref}}$ & 56492  \\
Position epoch ($\mathrm{MJD}$) & $T_{\mathrm{pos}}$ & 57205 \\
\hline
                \multicolumn{3}{c}{Fitted values} \\
                \hline
Right ascension & $\alpha$ & $ 3^\mathrm{h}37^\mathrm{m}43^\mathrm{s}.82703(92)_{-74}^{+72}$ \\
Declination & $\delta$ & $17^\circ15'14\farcs818(43)_{-35}^{+36}$ \\
Distance ($\mathrm{kpc}$) & $d$ & $1.3(75)_{-79}^{+85}$  \\
Right-ascension proper motion ($\mathrm{mas \, yr^{-1}}$) & $\mu_{\rm \alpha}$ & $5.(01)_{-18}^{+17}$  \\
Declination  proper motion ($\mathrm{mas\, yr^{-1}}$) & $\mu_{\rm \delta}$ & $-(0.85)_{-0.66}^{+0.69}$  \\
Radial proper motion ($\mathrm{mas\, yr^{-1}}$) & $\mu_{\rm d}$ & $4.(58)_{-30}^{+30}$ \\
Dispersion measure ($\mathrm{pc}\,\mathrm{cm}^{-3}$) & $\mathrm{DM}$ & $21.316(49)_{-15}^{+15}$ \\
Dispersion measure variation ($\mathrm{pc}\,\mathrm{cm}^{-3}\,\mathrm{yr}^{-1}$) & $\mathrm{DM}'$ & $(-6)_{-34}^{+34} \times 10^{-6}$
 \\
\hline
Rescaled spin frequency ($\si{Hz}$) & $\bar{f}$ & $365.9533379022(28)_{-25}^{+24}$  \\
Rescaled spin frequency derivative ($10^{-15}\, \si{Hz \, s^{-1}}$) & $\bar{f}'$ & $-2.35587(76)_{-69}^{+70}$ \\
\hline
 \multicolumn{3}{l}{\em Orbit of pulsar around centre of mass (CM) of inner binary} \\
Orbital period ($\mathrm{days}$) & $P_{\rm I}$ & $1.62940(06)_{-30}^{+30}$  \\
Projected semi-major axis (lt-s) & $a_{\rm p}\sin i_{\rm I}$ & $1.21752(80)_{-11}^{+11}$  \\
Inclination offset ($^{\circ}$) & $\delta i = i_{\rm I} - i_{\rm O}$ & $-0.00(29)_{-29}^{+29}$ \\
Laplace-Lagrange & $e_{\rm I}\sin \omega_{\rm I}$ & $6.93(65)_{-19}^{+19}\, \times \, 10^{-4}$ \\
Laplace-Lagrange & $e_{\rm I}\cos\omega_{\rm I}$ & $-8.5(44)_{-96}^{+95}\, \times \, 10^{-5}$ \\
Time of ascending node ($\mathrm{MJD}$) & ${t_{\mathrm{asc}}}_{\rm I}$ & $55917.15(84)_{-10}^{+10}$ \\
Long. of asc. nodes offset ($^\circ$) & $\delta\Omega = \Omega_{\rm O} - \Omega_{\rm I}$ & $0.00001(74)_{-34}^{+34}$ \\
\hline
 \multicolumn{3}{l}{\em Orbit of CM of inner binary around CM of the whole system} \\
Orbital period ($\mathrm{days}$) & $P_{\rm O}$ & $327.255(39)_{-51}^{+51}$ \\
Projected semi-major axis (lt-s) & $a_{\rm b}\sin i_{\rm O}$ & $74.6723(74)_{-58}^{+57}$ \\
Co-projected semi-major axis (lt-s) & $a_{\rm b}\cos i_{\rm O}$ & $91.4(35)_{-48}^{+47}$\\
Laplace-Lagrange & $e_{\rm O}\sin \omega_{\rm O}$ & $0.035114(31)_{-76}^{+76}$ \\
Laplace-Lagrange & $e_{\rm O}\cos\omega_{\rm O}$ & $-0.003524(80)_{-24}^{+24}$ \\
Time of ascending node ($\mathrm{MJD}$) & ${t_{\mathrm{asc}}}_{\rm O}$ & $56230.195(11)_{-41}^{+41}$ \\
Longitude of outer ascending node ($^{\circ}$) & $\Omega_{\rm O}$ & $-44.(34)_{-13}^{+14}$ \\
\hline
Inner mass ratio & $m_{\rm i}/m_{\rm p}$ & $0.1373(50)_{-18}^{+18}$  \\
SEP $\Delta$ & $\Delta$ & $(4.8)_{-9.4}^{+9.5} \times 10^{-7}$ \\
ToA uncertainty rescaling & $\mathrm{EFAC}$ & $1.31(53)_{-96}^{+94}$ \\
\hline
                \multicolumn{3}{c}{Derived values} \\
                \hline
Parallel proper motion ($\mathrm{km\, s^{-1}}$) & $V_{\parallel}$ & $29.(82)_{-93}^{+93}$  \\
Plane-of-sky proper motion ($\mathrm{km\, s^{-1}}$) & $V_{\bot}$ & $3(3.1)_{-2.5}^{+2.8}$  \\
\hline
Spin frequency ($\si{Hz}$) & $f$ & $365.9533630(00)_{-16}^{+16}$  \\
Spin frequency derivative ($10^{-15} \, \si{Hz \, s^{-1}}$) & $f'$ & $-2.32(44)_{-37}^{+41}$  \\
\hline
 \multicolumn{3}{l}{\em Orbit of pulsar around CM of inner binary} \\
Semi-major axis (lt-s) & $a_{\rm p}$ & $1.924(27)_{-61}^{+60}$  \\
Orbital inclination ($^{\circ}$) & $i_{\rm I}$ & $39.2(51)_{-14}^{+14}$ \\
Orbital eccentricity & $e_{\rm I}$ & $6.98(90)_{-31}^{+30}\, \times \, 10^{-4}$ \\
Longitude of periastron ($^{\circ}$) & $\omega_{\rm I}$ & $97.0(22)_{-75}^{+76}$ \\
Time of periastron passage ($\mathrm{MJD}$) & ${t_p}_{\rm I}$ & $55917.59(75)_{-14}^{+14}$ \\
Longitude of asc. node  ($^{\circ}$) & $\Omega_{\rm I}$ & $-44.(34)_{-13}^{+14}$ \\
\hline
 \multicolumn{3}{l}{\em Orbit of CM of inner binary around CM of the whole system} \\
Semi-major axis (lt-s) & $a_{\rm b}$ & $118.0(53)_{-37}^{+36}$ \\
Orbital inclination  ($^{\circ}$) & $i_{\rm O}$ & $39.2(37)_{-14}^{+14}$ \\
Orbital eccentricity & $e_{\rm O}$ & $0.035290(78)_{-78}^{+78}$ \\
Longitude of periastron  ($^{\circ}$) & $\omega_{\rm O}$ & $95.732(19)_{-27}^{+27}$  \\
Time of periastron passage ($\mathrm{MJD}$) & ${t_p}_{\rm O}$ & $56317.219(76)_{-52}^{+52}$ \\
\hline
Pulsar mass ($\Msol$) & $m_{\rm p}$ & $1.44(01)_{-15}^{+15}$  \\
Inner-companion mass ($\Msol$) & $m_{\rm i}$ & $0.197(80)_{-19}^{+19}$ \\
Outer-companion mass ($\Msol$) & $m_{\rm o}$ & $0.410(58)_{-40}^{+40}$ \\
\end{tabular}
\end{footnotesize}
\tablefoot{The error bars apply to the digits between parenthesis. Upper-case indices $\rm I,O$ refer to the inner and outer binary respectively while lower-case indices $\rm p,i,o$ refer to the pulsar, inner white dwarf and outer white dwarf respectively.}
\end{table*}

\begin{figure*}
\centering
\includegraphics[width=\textwidth]{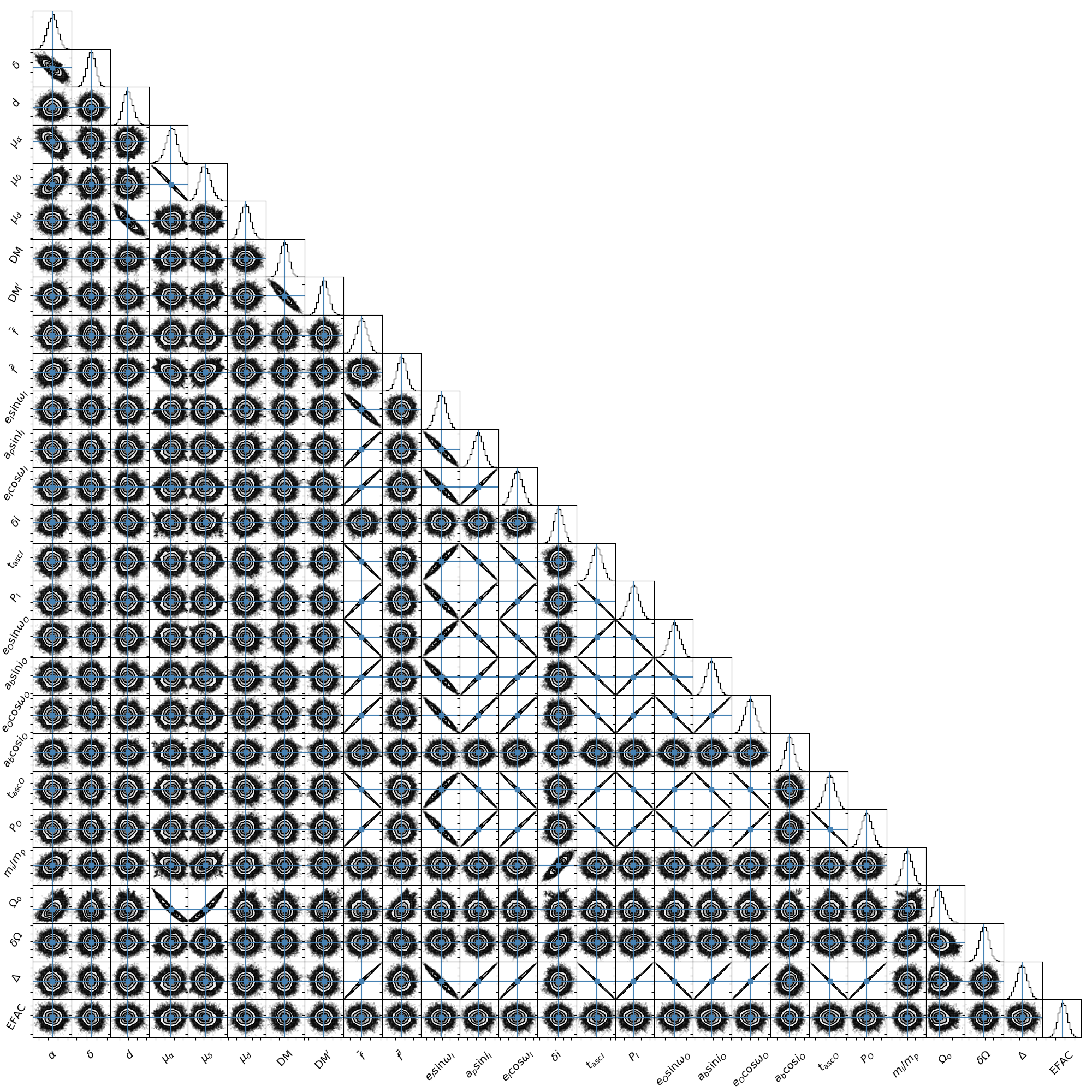}
\caption{Corner plot of the correlations between the fitted parameters of Table \ref{tab:fitresults} and their marginalised distribution (diagonal histograms) sampled by our MCMC. A high-resolution version is available as online supplementary material.\label{fig:corner}}
\end{figure*}
Our main goal in this work has been to get a Bayesian estimate of the uncertainties on each of the parameters of the problem, or in other words to estimate the posterior probability density function (PDF) of the parameters $\vec{\theta}$ belonging to model $M$ given our data $D$ using Bayes' rule,
\begin{equation}
	P(\vec{\theta}, M | D) = \frac{P(D |\vec{\theta}, M) P(\vec{\theta}, M)}{P(D)}.
\end{equation}
The prior function, $P(\vec{\theta}, M)$, was chosen flat except for astrometric parameters that benefited from prior knowledge of position and angular proper motion from the Gaia mission DR2 \citep{lindegren_gaia_2018}, of distance from photometric observations of the inner white dwarf \citep{ransom_millisecond_2014} and radial velocity from optical spectroscopy of the same star \citep{kaplan_spectroscopy_2014}. The Gaia DR2 catalogue does not model orbital motion which may then contaminate both position and proper motion. In the present case, given the distance of the source the magnitude of orbital motion is similar to the uncertainties reported by Gaia DR2. In order to account for potential systematic errors we have multiplied by two these uncertainties before using them as standard deviations of our Gaussian priors (see Table \ref{tab:priors}). We have also applied a factor of two to the uncertainty on the photometric distance reported in \citet{ransom_millisecond_2014} in order to account for potential systematic effects that would bias this measurement. For instance an inaccurate spectroscopic estimate of surface gravity (the `high log g problem' in low-mass white dwarfs, see \citet{tremblay_3d_2015} and references therein) would in turn bias the stellar radius estimate and therefore the absolute magnitude of the star. 
It is worth pointing out that the two commonly used free-electron density models for the Galaxy, NE2001 \citep{NE2001} and YMW16 \citep{YMW16}, both predict a distance of about 800 pc which is significantly smaller that reported in Tables \ref{tab:priors}-\ref{tab:fitresults}. This indicates that the electron density for the given Galactic height ($z=-690(40)\,$pc) is overestimated.
All priors are summarised in Table \ref{tab:priors}. Let us note that our fit for the radial proper motion $\mu_{\rm d}$ is unconstraining as the uncertainties reported in Table \ref{tab:fitresults} match the radial velocity prior of Table \ref{tab:priors}. The uncertainties of all the other fitted quantities are improved with respect to their prior.

The high dimensionality of the PDF  together with the necessity to integrate numerically the equations of motion makes the problem computationally challenging. However our C++ code is able to calculate one PDF value in less than 10\,s on a last-generation laptop, which made it possible to sample the PDF on a medium-size computer cluster. 
The sampling was achieved using a home-made implementation of the affine-invariant Markov-chain Monte Carlo (MCMC) of \cite{goodman_ensemble_2010} parallelised with the scheme of \cite{foreman-mackey_emcee_2013}. The advantage of this algorithm is to be efficient in high dimensionality \citep{allison_comparison_2014} and insensitive to any level of linear correlations between the parameters. This is particularly important as we found $\Delta$ to be highly correlated with many orbital parameters (see Figure \ref{fig:corner}). However, we also found that non-linear "correlations" between parameters were preventing convergence within a reasonable time, which was solved by appropriate re-parametrisation (see Section \ref{sec:param}). Convergence was evaluated by requiring that fluctuations of the mean and standard-deviation estimators be smaller that $6\%$ of the full-chain standard deviation for each parameter (see e.g. \cite{dunkley_fast_2005}, Section \ref{sec:mcmcconvergence} and chain plot in supplementary online material). We noticed that standard deviations sometimes converge later than means, particularly for $\Delta$, confirming the importance of monitoring both indicators to ensure reliable uncertainties.


Due to its very low ecliptic latitude, $\sim2\,\deg$, the timing of PSR J0337+1715 is potentially sensitive to a range of effects occurring in the Solar-system. In particular, we detected in preliminary runs a slight increase in timing residuals of the order of $1\,\mathrm{\mu s}$ when the pulsar was within $3\deg$ of the Sun. We attributed this increase to the inaccuracy of the Solar-wind electron density model used by \textsc{tempo2} to calculate the related DM. We mitigated this effect by removing all the ToAs taken within $5 \deg$ of the Sun. Moreover, our periodogram shows a secondary $\sim 0.2\, \mathrm{\mu s}$ component close to the Earth orbital frequency, sign of possible extra inaccuracies in the Solar-wind model or in the Solar-system ephemerides. This is likely to affect outer-orbit parameters since this period is close to 1 year but such a correlation can only widen posterior uncertainties.

\begin{figure}[htb]
	\centering
	\includegraphics[width=\columnwidth]{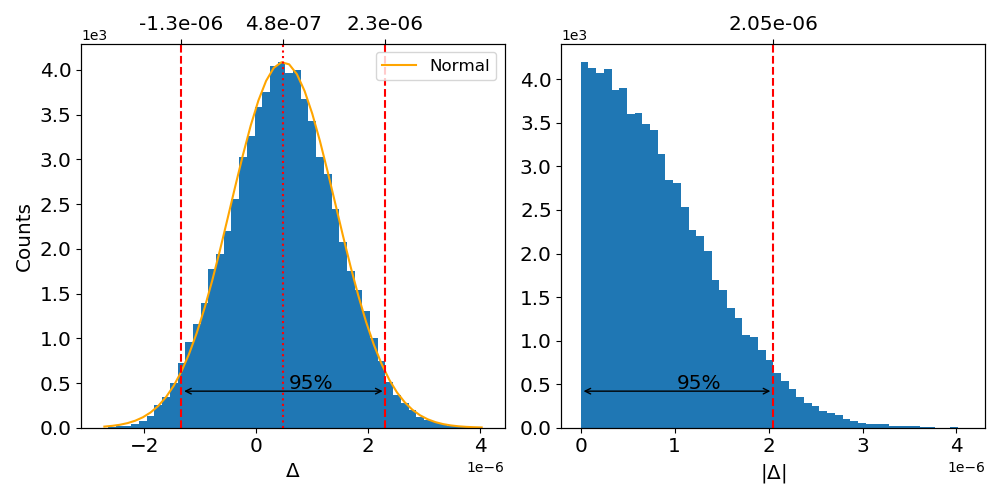}
	\caption{Left-hand side: Marginalised posterior probability distribution of the SEP violation parameter $\Delta$ sampled by MCMC and normal law with the same mean and standard deviation. The upper axis gives the mean value and the boundaries of the $95\%$ confidence region. Right-hand side: distribution of distance to GR derived from the left-hand-side distribution. }
	\label{fig:SEPDdistro}
\end{figure}

Two models $M$ were tested. Our main model includes only $\Delta$ as a free parameter while our secondary model includes the three additional 1PN-strong-field parameters, $\bar{\beta}_{\rm p}$, $\bar{\beta}_{\rm pp}$, $\bar{\beta}^{\rm p}$, yielding the following 95\% C. L. constraints for them:
\begin{eqnarray}
    \Delta &=& 1.1_{-4.6}^{+4.8} \times 10^{-6} \label{eq:DeltaM2}\,,\\
    \bar\beta_{\rm p}  &=& -2_{-14}^{+13} \label{eq:blim1}\,,\\
    \bar\beta_{\rm pp} &=& -0.1_{-1.0}^{+0.9} \label{eq:blim2}\,, \\
    \bar\beta^{\rm p}  &=& +0.8_{-6.6}^{+7.2} \label{eq:blim3}\, .
\end{eqnarray}
 In regard to binary-pulsar tests (see Section~\ref{sec:tests_theory}), the above results on the three $\bar{\beta}$ parameters are unconstraining. We used this prior knowledge to run our main model with $\bar{\beta}_{\rm p} = \bar{\beta}^{\rm p} = \bar{\beta}_{\rm pp} = 0$ and obtain our primary SEP limit
\begin{equation} \label{eq:limit}
  \Delta = (+0.5 \pm 1.8) \times 10^{-6} \quad\mbox{(95\% C.L.)},
\end{equation}
which translates into $|\Delta| < 2.05\, \times \, 10^{-6}$ at $95\%$ C. L. (see Figure \ref{fig:SEPDdistro}). The full result of the main model is reported in Table \ref{tab:fitresults}. Note that $\sim 8\%$ of the reported uncertainties are due to unaccounted systematics absorbed in the EFAC parameter (see also Section \ref{sec:residualanalysis}). The wider uncertainty obtained in the secondary run is due to large correlations with the three additional parameters.


\subsection{MCMC run and convergence \label{sec:mcmcconvergence}}

The affine-invariant algorithm of \cite{goodman_ensemble_2010} requires to move $N$ walkers together at each iteration. The gist of this algorithm is that the walkers within the set are not independent from each other while the set as a whole constitutes a single effective walker in the Markov process sense, namely that it depends only on its previous state. Individual moves within the set are informed by the positions of other walkers in a way that renders the algorithm rigorously immune to any linear correlation, or any affine parameter transformation. However it might be sensitive to correlation of a higher degree, or to non-convexity of the posterior isosurfaces. Therefore, with this algorithm one should take care of removing as much as possible any non-linear correlations by choosing an appropriate parameter set (see Section \ref{sec:param}) but very large linear correlations, as can be seen in Figure \ref{fig:corner}, are well resolved by the algorithm.

The authors of \cite{goodman_ensemble_2010, foreman-mackey_emcee_2013} recommend to choose a number of walkers within the set much larger than the number of parameters. In the present case we chose to use 288 walkers per set. The only other tunnable parameter is the unique parameter of the proposal function, $a$, which controls the size of the steps that can be attempted. The authors of \cite{goodman_ensemble_2010, foreman-mackey_emcee_2013} suggest the value $a=2$ in order to keep an acceptance fraction $\sim 0.4$. We found that, as the chain was getting closer to convergence the acceptance fraction could drop dramatically, sign of non-linear correlations or non-convexity. This drop was largely mitigated by adopting the final parameter set of Section \ref{sec:param}, but we still had to choose a smaller step-size parameter $a=1.6$ to keep the acceptance fraction close to the level mentioned above. 

Due to the large parameter space and the computing time needed to compute one $\chi^2$ (about 10\,s) 
we parallelised the MCMC code using the scheme of \citet{foreman-mackey_emcee_2013}. This allowed us to run the MCMC using 144 cores of the meso-scale MesoPSL cluster (see acknowledgements) each calculating 2 walkers (which is the minimum number of walkers per core given the algorithm used). A few 10,000 steps were typically necessary to reach convergence. However, 
it is important to quantitatively estimate convergence as one cannot afford to run the MCMC for an unnecessarily large number of iterations. 
Beyond visual inspection of the parameter chains which allows to discard any obvious burn-in phase, we also monitored the autocorrelation time of each chain \citep{goodman_ensemble_2010}. However, we found that the most constraining convergence estimator was to monitor the variation of the mean and standard deviation of each parameter chain. Formally, one needs to compute the value of the following estimator on each chain \citep{dunkley_fast_2005}, 
\begin{equation}
    \hat{r}_{\hat{o} = \hat{m}, \hat{\sigma}} = \frac{\hat{\sigma}\left(\hat{o}\right)}{\sigma},
\end{equation}
where $\hat{\sigma}$ is a standard-deviation estimator, $\hat{o}$ is a statistical estimator which here is either the mean $\hat{m}$ or the standard deviation $\hat{\sigma}$, and $\sigma$ is the standard deviation estimated on the entire length of the chain.
In practice, we recorded the 288 walkers every 5 iterations and use the last 69408 recorded elements (241 independent sets of 288 walkers). The standard deviation of $\hat{o}$ was estimated by applying $\hat{o}$ on 20 sub-samples of the chain and then estimating the standard deviation of the set of the values obtained. The chain was considered converged if both $\hat{r}_{\hat{m}}$ and $\hat{r}_{\hat{\sigma}}$ estimated values are smaller than $0.06$. A situation we observed is when the latter keeps varying significantly while the former is stable and under the cutoff. In other words, the chain widens with constant mean, rendering a mean-based convergence estimator uninformative and possibly leading to an underestimation of the parameter uncertainties.


\subsection{Analysis of the residuals \label{sec:residualanalysis}}

\begin{figure*}
    \centering
    \includegraphics[width=0.99\textwidth]{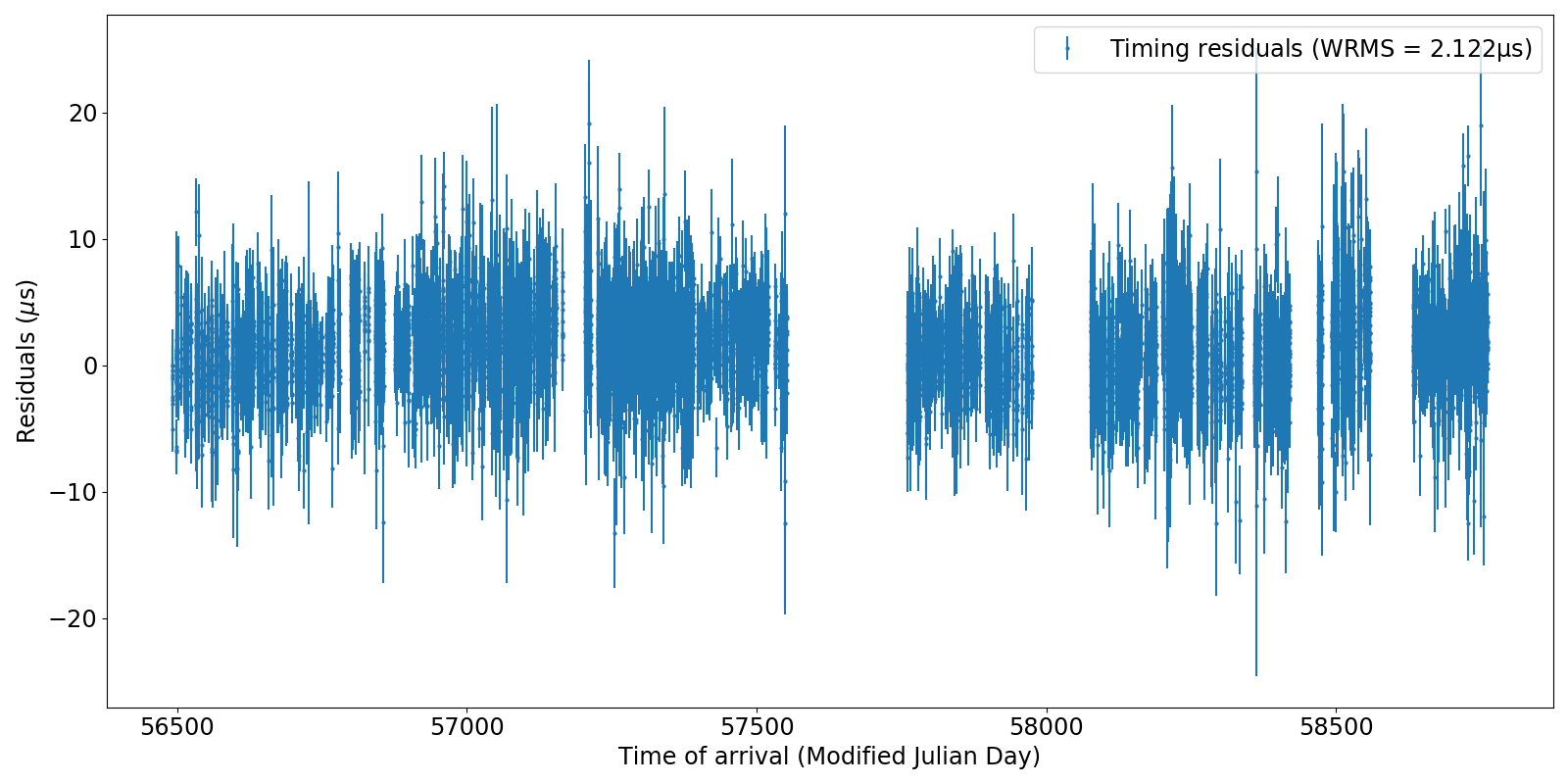}
    \caption{Residuals of the times of arrival using the mean parameters reported in Table \ref{tab:fitresults}.}
    \label{fig:residuals}
\end{figure*}

\begin{figure*}
    \centering
    \includegraphics[width=0.99\textwidth]{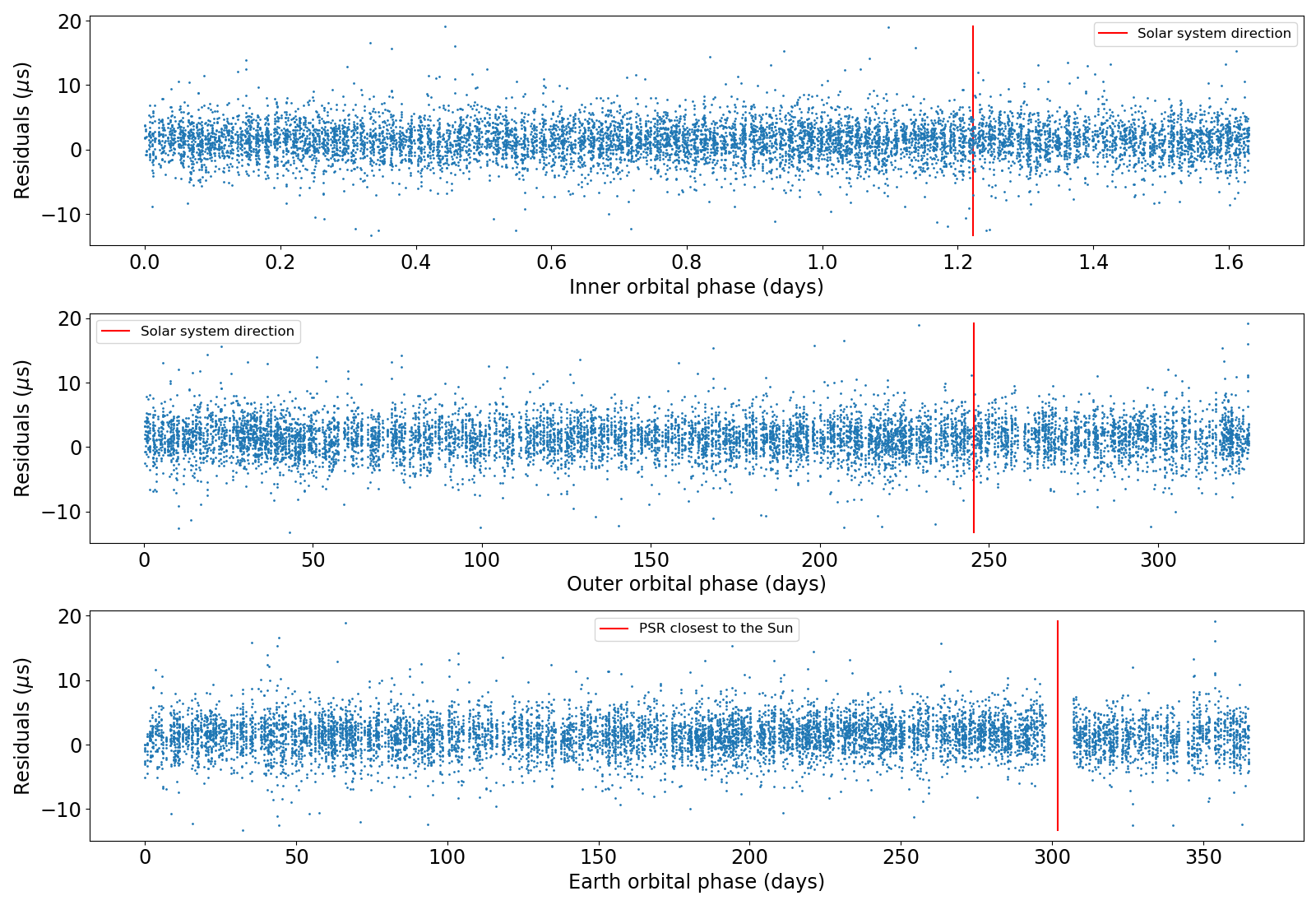}
    \caption{Residuals of the times of arrival corresponding to the parameters of Table \ref{tab:fitresults} plotted versus the inner-binary orbital phase (top), the outer-binary orbital phase (middle) and the Earth orbital phase (bottom). The red vertical bars of the first two plots show the phase where the pulsar (resp. the inner binary) is closer to the Solar System. The red vertical bar on the bottom plot, shows the Earth orbital phase where the line of sight to the pulsar passes closest to the Sun. We have removed every ToA when the pulsar is less than $5\deg$ from the Sun, which explains the gap around that particular phase. }
    \label{fig:resorbphase}
\end{figure*}

We have assessed the robustness of our fit by evaluating the presence of systematic components in the residuals (Figure \ref{fig:residuals}). As it appears from Figure \ref{fig:resorbphase}, no significant structure is present at either the inner or outer orbital period, nor at the Earth orbital period notwithstanding the sharp cut around the phase of closest approach to the Sun made to prevent any bias induced by unmodelled DM contributions.

\begin{figure*}
     \centering
     \includegraphics[width=\textwidth]{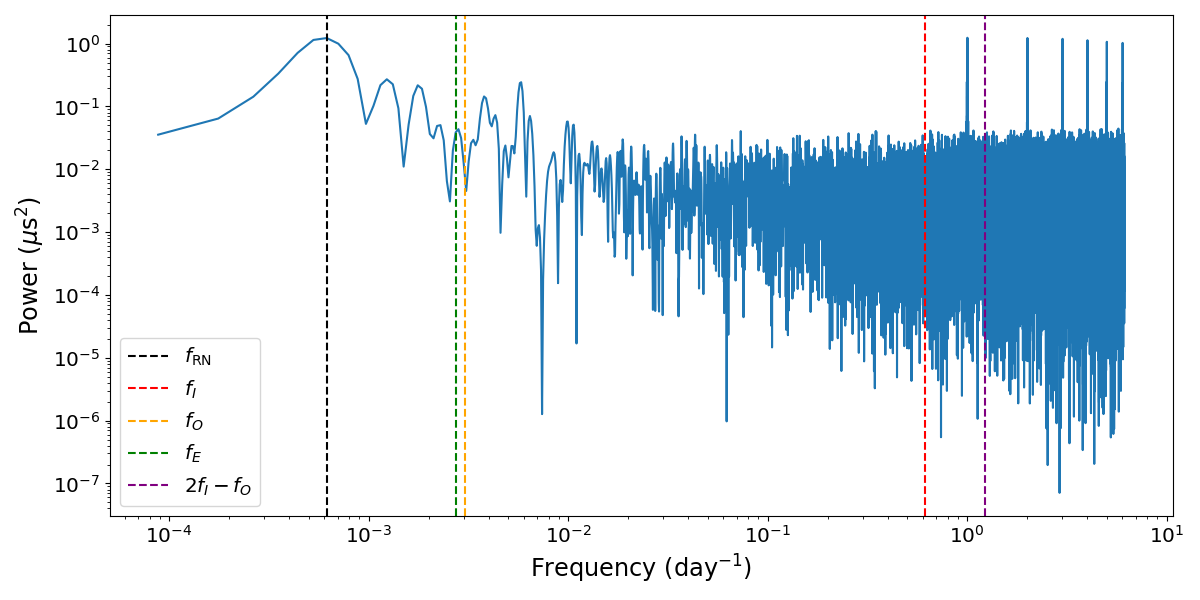}
     \caption{Lomb-Scargle periodogram of the timing residuals for the solution of Table \ref{tab:fitresults}. The periodogram is sampled at $f_m/5$ where $f_m \simeq 2268\mathrm{days}^{-1}$ is the inverse of the full time span. From left to right vertical lines show the frequencies of the red-noise component ($f_{RN}\simeq 1650\mathrm{day}^{-1}$, black), the Earth orbital period ($f_E$, green), the outer-orbit orbital period ($f_O$, orange), the inner-binary orbital period ($f_I$, red) and the SEP violation signature ($2f_I - f_O$, purple).}
     \label{fig:periodogram}
 \end{figure*}

In order to estimate more thoroughly the presence of systematic modulations we produced a Lomb-Scargle periodogram \citep{lomb_least-squares_1976, scargle_studies_1982} of the same residuals, Figure \ref{fig:periodogram}. It appears that indeed there is no significant power at the frequencies mentioned above, except for a tentative signal $\lesssim 0.2\, \mathrm{\mu s}$ at the Earth orbital frequency. Because of the proximity of this frequency with the outer-binary orbital frequency this might lead to correlate Solar-system and outer-binary effects and therefore enlarge uncertainties related to the parameters involved. 

The dominant component is the low-frequency peak and its subsequent harmonics which we interpret as time-correlated red noise. A number of causes have been proposed in the literature. The main ones are the intrinsic spin frequency noise \citep{shannon_assessing_2010, melatos_pulsar_2014} or magnetospheric fluctuations \citep{lyne_switched_2010}. It has also been proposed that asteroid belts could result in apparent timing noise \citep{shannon_asteroid_2013}. Propagation effects under the form of long-term variations of the dispersion measure along the line of sight due to turbulence in the interstellar medium could be the cause in some cases \citep{keith_measurement_2013}. We have tried to split the time span of our observations into several intervals with different DM values, but the fit was consistent with an absence of variation discarding this explanation. Red noise can also have a local cause, such as irregularities of the terrestrial time realisation used to time the pulsar \citep{hobbs_pulsar-based_2020,hobbs_development_2012} or inaccuracies in the planetary masses used to generate the Solar System ephemeris \citep{champion_measuring_2010, caballero_studying_2018}. However, we use the 2016 realisation of the BIPM terrestrial time which \cite{hobbs_development_2012} has confirmed as suitable for precision pulsar timing. Besides, if the red noise was caused by any inaccuracy in planetary masses, then the signature would be at the orbital frequency of the responsible planet \citep{champion_measuring_2010}. The only orbital period in the Solar System that approaches the $1650$ days of the red-noise signal is the orbital period of the dwarf planet Ceres. However the uncertainty on its mass in the NASA JPL ephemeris DE430 \citep{folkner_planetary_2014} we use in this work is too small to explain a signal of that magnitude. 

Thus, there is no deterministic model that can be fitted to that component, but since its frequency is much lower than any other in the system it is unlikely to bias the parameters. However, it results in increasing the scale of the ToA uncertainties (EFAC parameter in Table \ref{tab:fitresults}) in order to accommodate this systematic effect into a reduced $\chi^2$ equal to unity. Were the PDF perfectly Gaussian, that would result into posterior uncertainties increased in exactly the same proportion, which we can here estimate at $\sim 8 \%$. Therefore, our posterior uncertainties should be seen as upper limits. Interestingly, because the analysis of \cite{archibald_universality_2018} focuses on a specific frequency signature for the SEP, the result quoted in that work should be seen as a lower limit on the actual uncertainty in the sense of the above discussion. 

Finally, the prominent peak at $1\, \mathrm{day}^{-1}$ and its harmonics on the periodogram of Figure \ref{fig:periodogram} result from the convolution of the red noise component with the observing window functions of our data. Indeed, due to its meridian configuration, the Nan\c{c}ay radio telescope can only observe the same object during $\sim 1$h windows separated by an integer number of sidereal day. Due to the proximity of the inner orbital frequency with the observing frequency of $\sim 1 \mathrm{day}^{-1}$ one might be concerned with a potential bias. However we have checked that the Fourier transform of a comb of 1h window functions results in sharp narrow peaks whose width does not exceed a few percents of the daily frequency and therefore cannot significantly bias signals at the orbital frequencies.

\section{Theory dependent tests\label{sec:tests_theory}}

The parametrised post-Newtonian (PPN) formalism \citep[see e.g.][]{Will_book}, with its ten theory-independent parameters, has proven to be a powerful tool to quantify and compare tests of GR and its alternatives in the weak-field environment of the Solar System. Unfortunately, there is no such framework that generically extends beyond the weak field approximation of the PPN formalism and therefore is able to cover gravity experiments with strongly self-gravitating bodies, like the one in this paper. One reason is that, unlike in the Solar System, the treatment of the motion of a neutron star in an external gravitational field requires the full complexity of a gravity theory \citep{Will_2018}. To put the UFF test conducted in this paper into context with other experiments, in particular Solar System tests and gravitational wave tests with binary pulsars, it has been suggested to use theory-dependent frameworks \citep[see e.g.][]{Damour_2009}. Scalar-tensor theories of gravity have turned out to be particularly useful for this purpose. Apart from being well motivated and well studied alternatives to GR \citep{fujii2007scalar}, they show a rich phenomenology in their deviations from GR, including prominent effects related to the non-linear strong-field regime of neutron stars \citep[see e.g.][]{DEF_1993, damour_tensor-scalar_1996}. 

In this paper, as a theory-dependent framework we use the class of Bergmann-Wagoner theories. Bergmann-Wagoner theories represent the most general scalar–tensor theories with one scalar field that are at most quadratic in the derivatives of the fields \citep{Will_book}. Quite a number of well known scalar-tensor theories belong to this class, like Jordan-Fierz-Brans-Dicke (JFBD) gravity \citep{jordan1955schwerkraft, Fierz_1956,Brans_1961}, DEF gravity \citep{DEF_1993}, MO gravity \citep{Mendes_2016}, $f(R)$ gravity \citep{LRR_fR}, and massive Brans-Dicke gravity \citep{Alsing_2012}. Bergmann-Wagoner theories form a subclass of the class of Horndeski theories \citep{Horndeski_1974}, which is the most general class of mono-scalar-tensor theories in four dimensions whose Lagrangian leads to second order field equations.

In the following we interpret our limits of Section~\ref{sec:parameters} in two different approaches within the class of Bergmann-Wagoner theories. In the first approach we remain (mostly) generic, in the sense that make as few assumptions as possible concerning the details of the theory. In the second approach, we pick a specific two-parameter scalar-tensor theory, that is,\ DEF gravity. For this two-parameter class of theories we can then explicitly calculate the properties of  neutron stars for different equations of state (EOS) and derive constraints on the two-dimensional theory space from the limits presented here.


\subsection{Generic tests within Bergmann-Wagoner scalar-tensor gravity}

In Bergmann-Wagoner theories, the field equations for the (physical) metric $g_{\mu\nu}$ and the scalar field $\phi$ are a result of the following action
\begin{eqnarray}
  S &=& \frac{1}{16\pi G_0}\int\sqrt{-g}\,d^4x 
    \left(\phi R - \frac{\omega(\phi)}{\phi} \partial_\mu\phi\partial^\mu\phi - U(\phi)\right) 
    \nonumber\\
    && + S_{\rm mat}\left[\psi_{\rm mat}^A,g_{\mu\nu}\right] \,,   
\end{eqnarray}
where $G_0$ is the fundamental (`bare') gravitational constant, $g$ the determinant of $g_{\mu\nu}$, $R$ the curvature scalar, $\omega(\phi)$ is the coupling function, and $U(\phi)$ the scalar potential. The physical (Newtonian) gravitational constant, as measured in a Cavendish-type experiment, is given by
\begin{equation}
    G_{\rm N} = \frac{G_0}{\phi_0(1 - \zeta)} \,,
\end{equation}
where $\phi_0$ denotes the cosmological background field and $\zeta \equiv 1/(2\omega(\phi_0) + 4)$. $S_{\rm mat}$ is the action of the matter fields $\psi_{\rm mat}^A$, which couple universally to the spacetime metric $g_{\mu\nu}$. For our discussion, we assume that $U(\phi)$ can be neglected on the scale of the triple system, that is,\ $U''(\phi) \ll 1/a_b^2$. In terms of a massive scalar field, this means that we assume that the Compton wavelength is much larger than the extension of the system  \citep[see][on how J0337+1715 can be used to constrain massive scalar fields]{Seymour_2019}.

The effective gravitational constant that enters the $N$-body Lagrangian is given by
\begin{equation} \label{eq:Gsasb}
    G_{ab} = G_{\rm N}\left[1 - 2 \zeta(s_a + s_b - 2s_a s_b)\right],
\end{equation}
where the sensitivity
\begin{equation}
    s_a \equiv \left(\frac{d\ln m_a(\phi)}{d\ln\phi}\right)_{\phi_0}
\end{equation}
accounts for the dependence of each body on a change in the ambient scalar field, while the number of baryons remains fixed \citep{Will_book}. For neutron stars, $s_a$ depends on the EOS. It is typically of the order of 0.1 but, depending on the details of $\omega(\phi)$, its (absolute) value can be very large, as we discuss further below. For the {\em relative GWEP parameter} one finds
\begin{equation}
    \Delta_{ab} = -2 \zeta (s_a + s_b - 2s_a s_b).
\end{equation}
Because of the product $s_a s_b$, in general it is not possible to interpret the quasi-Newtonian equations of motion in terms of inertial and gravitational masses of the individual bodies in an $N$-body system \citep{Will_book}. For weakly self-gravitating bodies the sensitivity $s_a$ is simply related to the fractional gravitational binding energy $ \varepsilon_{{\rm grav},a}$ via
\begin{equation}\label{eq:sa_weak_field}
    s_a = -(1 + 2\lambda) \, \varepsilon_{{\rm grav},a} 
          + {\cal O}\left(\varepsilon_{{\rm grav}, a}^2\right)\, ,
\end{equation}
where $\lambda \equiv \phi_0\omega'(\phi_0)\zeta^2(1 - \zeta)^{-1}$. The two Eddington parameters of the PPN formalism are given by
\begin{equation}
    \gamma_{\rm PPN} = 1 - 2\zeta \,,\quad
    \beta_{\rm PPN}  = 1 + \zeta\lambda \,,
\end{equation}
and the Nordtvedt parameter of equation~(\ref{eq:Delta_a}) reads 
\begin{equation} \label{eq:etaBW}
    \eta = 4\beta_{\rm PPN} - \gamma_{\rm PPN} -3 = 2\zeta(1 + 2\lambda) \,.
\end{equation}
For the inner and outer white dwarf we have $ \varepsilon_{\rm grav,i} \simeq -1.8 \times 10^{-5}$ and $\varepsilon_{\rm grav,o} \simeq -4.8 \times 10^{-5}$, respectively. Consequently $G_{\rm io} \simeq G_{\rm N}$ for the interaction between the two white dwarfs, and $G_{{\rm pi}} \simeq G_{{\rm po}} \simeq G_{\rm N}(1 - 2 \zeta s_{\rm p})$ for the interaction between the pulsar and the white dwarfs. Hence, our result for $\Delta$ in Table~(\ref{tab:fitresults}) leads to a direct constraint for 
\begin{equation} \label{eq:zsplimit}
    \zeta s_{\rm p} \simeq -\frac{\Delta}{2}
        = (-0.2 \pm 0.9) \times 10^{-6} \quad\mbox{(95\% C.L.)}\, ,
\end{equation}
where $\Delta \equiv \Delta_{{\rm p}b}(s_b=0)$.  The above limit can be considered as generic within the family of Bergmann-Wagoner theories of gravity, in the sense that it does not require a specification of the coupling function $\omega(\phi)$. Later, we use this limit to impose constraints on the parameter space of a specific two parameter family of Bergmann-Wagoner theories. Before that, we need to discuss the strong-field modifications at the post-Newtonian level of the 3-body dynamics.


\subsubsection*{First post-Newtonian contributions}

At the first post-Newtonian (1PN) level (order $v^2/c^2$), the PPN parameters $\bar\beta_{\rm PPN} \equiv \beta_{\rm PPN} - 1$ and $\bar\gamma_{\rm PPN} \equiv \gamma_{\rm PPN} - 1$ need to be replaced by the body-dependent quantities $\bar\gamma_{ab}$ and $\bar\beta^a_{bc}$ (see Appendix~\ref{sec:eqofmotion} for details). These strong-field generalisations of the PPN Eddington parameters depend on the sensitivities and their derivatives of the bodies in a system. For the detailed expressions, we refer the reader to \cite{will_theory_1993} and \cite{DEF_1992}. The latter uses the so-called Einstein-frame representation and gives these terms for multi-scalar-tensor theories. More generally, in the triple system of PSR~J0337+1715, where two of the bodies are weakly self gravitating, one finds for the twelve 1PN-strong-field parameters, to good approximation,
\begin{eqnarray}
    \bar\gamma_{\rm io}     &\simeq& \bar\gamma_{\rm PPN} \,,\\
    \bar\gamma_{\rm pi} &\simeq& \bar\gamma_{\rm po} 
        \;\;\simeq\;\; \bar\gamma_{{\rm p}b}(s_b = 0) \equiv \bar\gamma_{\rm p}\,,\\
    \bar\beta^{\rm i}_{\rm oo} &\simeq& \bar\beta^{\rm o}_{\rm ii} \;\;\simeq\;\; \bar\beta_{\rm PPN} \,,\\
    \bar\beta^{\rm i}_{\rm po} &\simeq& \bar\beta^{\rm o}_{\rm pi} 
        \;\;\simeq\;\; \bar\beta^a_{{\rm p}c}(s_a=s_c=0) \equiv \bar\beta_{\rm p} \,,\\
    \bar\beta^{\rm i}_{\rm pp} &\simeq& \bar\beta^{\rm o}_{\rm pp} 
        \;\;\simeq\;\; \bar\beta^{a}_{\rm pp}(s_a=0) \equiv \bar\beta_{\rm pp} \,,\\
    \bar\beta^{\rm p}_{\rm ii} &\simeq& \bar\beta^{\rm p}_{\rm io}  
        \;\;\simeq\;\; \bar\beta^{\rm p}_{\rm oo} 
        \;\;\simeq\;\; \bar\beta^{\rm p}_{bc}(s_b=s_c=0) \equiv \bar\beta^{\rm p} \,,
\end{eqnarray}
leaving us with six different parameters at the 1PN level of the modified Einstein-Infeld-Hoffmann equations of motion, instead of the two in the weak field limit. Note the following symmetries: $\bar\gamma_{ab} = \bar\gamma_{ba}$ and $\bar\beta^a_{bc} = \bar\beta^a_{cb}$. 

At this stage, we can further reduce the number of 1PN parameters, without making more detailed assumptions about the theory, for instance about $\omega(\phi)$. The tight limits on $\bar\beta_{\rm PPN}$ and $\bar\gamma_{\rm PPN}$ from Solar System tests $\sim 10^{-5}$ \citep{Will_book}, directly put tight constraints on two of the six 1PN parameters. Furthermore, $\bar\gamma_{\rm p}$ only depends on terms proportional to $\zeta$ and $\zeta s_{\rm p}$, the first being constrained to $\sim 10^{-5}$ by Cassini \citep{Bertotti_2003} and the latter to $\sim 10^{-6}$ already by the Newtonian-level dynamics of the triple system (cf.\ equation~(\ref{eq:zsplimit}); see also limit~(\ref{eq:DeltaM2})). Consequently, without loss of generality, $\bar\gamma_{ab}$, $\bar\beta^{\rm i}_{\rm oo}$, and $\bar\beta^{\rm o}_{\rm ii}$ can be ignored in a self-consistent gravity test with the PSR~J0337+1715 system. Besides the $\Delta$ at the Newtonian level, we are left with the 1PN strong-field parameters $\bar\beta_{\rm p}$, $\bar\beta_{\rm pp}$, and $\bar\beta^{\rm p}$. These three parameters cannot be constrained without further assumptions, as we discuss below. Hence we have implemented a model based on deviations from GR parametrised by $(\Delta, \bar\beta_{\rm p}, \bar\beta_{\rm pp}, \bar\beta^{\rm p})$, which is called {\em secondary model} in Section~\ref{sec:numerical_model} and \ref{sec:parameters}. Our analysis based on this model leads to the generic limits (\ref{eq:DeltaM2}) and (\ref{eq:blim1}) -- (\ref{eq:blim3}).

In our generic approach, the parameters $\bar\beta_{\rm p}$, $\bar\beta_{\rm pp}$ can a-priori only be constrained if we make further assumptions and apply existing constraints from binary-pulsar systems. The reason is as follows. The three $\bar\beta$ parameters have terms, which are proportional to $\lambda\zeta$, $\zeta^2 s_{\rm p}^2$, $\lambda\zeta s_{\rm p}$, $\lambda\zeta s_{\rm p}^2$, and $\zeta^2 s_{\rm p}'$, where
\begin{equation}
     s_a' \equiv \left(\frac{d^2\ln m_a(\phi)}{d(\ln\phi)^2}\right)_{\phi_0} 
\end{equation}
\citep[see e.g.][for details]{Will_2018}. Solar System constraints on $\beta_{\rm PPN}$ and $\eta$ put tight constraints ($\sim 10^{-5}$) on $\lambda\zeta$, and $\zeta^2 s_{\rm p}^2$ is constrained to $\sim 10^{-12}$ because of equation~(\ref{eq:zsplimit}). However, the quantities $\lambda$, $s_p$, and $s_p'$ are  unconstrained by above considerations. Consequently, $\lambda\zeta s_{\rm p}$, $\lambda\zeta s_{\rm p}^2$, and $\zeta^2 s_{\rm p}'$ are a-priori unconstrained. Below, in a more theory specific context, we discuss a situation in DEF gravity where $\zeta s_{\rm p}^2$ can be of order unity, even if $\zeta s_{\rm p} \ll 1$ (spontaneous scalarisation of neutron stars). In such a highly non-linear strong field regime, $s_{\rm p}$ and $s_{\rm p}'$ can assume very large values. As a consequence, the $\bar\beta$ terms can become much larger than $\beta_{\rm PPN}$, even under condition (\ref{eq:zsplimit}). To give an example, in the regime of spontaneous scalarisation in DEF gravity, one finds $\bar\beta_{\rm pp}/\bar\beta_{\rm PPN} \sim \zeta^{-1}$ while $\bar\beta_{\rm pp}$ remains practically unaffected when $\zeta \rightarrow 0$ \citep[cf.][]{damour_tensor-scalar_1996}.

If we restrict $\lambda$ to values which are not very large, it can be shown that $\bar\beta_{\rm p}$ can be considered as small as well, since it only contains $\lambda\zeta s_{\rm p}$ as an a priori unconstrained term, where $\zeta s_{\rm p}$ has to be small according to equation~(\ref{eq:zsplimit}). 

Further constraints can come from binary pulsar experiments. In particular, dipolar-radiation tests in binary pulsars with spectroscopic white dwarfs \cite[see e.g.][]{Lazaridis_2009, freire_1738+0333}, can in principle provide generic constraints on $\zeta s_{\rm p}^2$, although some additional assumptions are needed, for instance for $U(\phi)$ \citep[cf.][]{Alsing_2012}. If $U(\phi) = 0$, the change in the orbital period of a pulsar-white dwarf binary, $P_{\rm b}$, due to dipolar radiation damping is given by 
\begin{equation} \label{eq:DGW}
    \dot{P}_{\rm b}^{\rm dipole} \simeq
       -\frac{G_{\rm N}}{c^3} \,
       \frac{16\pi^2}{P_{\rm b}} \,
       \frac{m_{\rm p} m_{\rm c} }{m_{\rm p} + m_{\rm c}} \,
       \frac{1 + e^2/2}{(1 - e^2)^{5/2}} \,
       \zeta s_{\rm p}^2
\end{equation}
\citep[see e.g. equation~(12.32) in ][]{Will_book}, where $e$ denotes the orbital eccentricity, and $m_{\rm p}$ and $m_{\rm c}$ are the masses of pulsar and white-dwarf companion respectively. To apply constraints from other pulsar to the PSR~J0337+1715 system it also requires some on the dependence of the sensitivity of the pulsar, $s_{\rm p}$, on the pulsar mass. In the strong-field regime of neutron stars this dependence can be highly non-linear \citep{damour_tensor-scalar_1996,Shao_2017}. PSR~J1738+0333 is a pulsar with a mass similar to PSR~J0337+1715 ($m_{\rm p} \simeq 1.46\,M_\odot$). The dipolar radiation test by \cite{freire_1738+0333,Zhu_1713} leads to
\begin{equation}
    \zeta s_{\rm p}^2 = (-0.5 \pm 1.7) \times 10^{-6} \quad\mbox{(95\% C.L.)}\,. 
\end{equation}
As a result, $\bar\beta_{\rm pp}$ can, in general, also be assumed to be small in the PSR~J0337+1715 system. 

Imposing a generic constraint on $\zeta^2 s_{\rm p}'$, and therefore on $\bar\beta^{\rm p}$, is somewhat less direct. $s_{\rm p}'$ enters, for instance, the precession of periastron, $\dot\omega$, which is particularly well tested -- in combination with other post-Keplerian parameters -- in eccentric short-orbital-period binary pulsars \citep{Wex_2014,Will_book}. However, only the so called Double Pulsar allows for a generic constraint on deviations from GR in $\dot\omega$, which is of the order of $10^{-3}$ \citep{Kramer_2009}. However, the masses of the Double Pulsar are significantly lower than the mass of PSR~J0337+1715. Nevertheless, the general agreement of all these systems with GR at least suggests that $\zeta^2 s_{\rm p}'$, and therefore $\bar\beta^{\rm p}$ can generally be assumed to be small as well. Moreover, given that the Cassini experiment already imposes $\zeta^2 \lesssim 10^{-10}$, $s_{\rm p}'$ would have to assume quite extreme values to lead to a significant $\bar\beta^{\rm p}$, certainly in view of the (still) quite weak limit (\ref{eq:blim3}).

To summarise, under additional assumptions, which we consider as reasonable for most situations, all three strong-field $\bar\beta$ parameters are tightly constrained by a combination of equation~(\ref{eq:zsplimit}) with constraints from binary pulsars experiments. The limits in equations~(\ref{eq:blim1}) to (\ref{eq:blim3}) are therefore generally not of particular interest, at least for constraining the class of scalar-tensor theories considered in this section.


\subsection{EOS-agnostic constraints on Damour--Esposito-Far{\`e}se (DEF) gravity}

In order to explicitly calculate $s_{\rm p}$ and $s_{\rm p}'$ and therefore $\Delta$ and the 1PN strong-field parameters that enter our equations of motion, one has to pick a specific theory of gravity, which we do in this subsection. In the quadratic mono-scalar tensor theory $T_1(\alpha_0,\beta_0)$ of \cite{DEF_1993}, the coupling function in the Einstein frame is quadratic in the scalar field, meaning that the coupling strength between the scalar field and the trace of the stress-energy tensor becomes field dependent in a linear way. In the Jordan-frame representation with the physical metric $g_{\mu\nu}$, which we are using here, the coupling function then reads
\begin{equation}
    \omega(\phi) = \frac{1}{2}\left(\frac{1}{\alpha_0^2 - \beta_0\ln\phi} - 3\right),
\end{equation}
where $\phi_0 \equiv 1$, without loss of generality \citep{Will_book}. Furthermore, one finds
\begin{equation} \label{eq:zlDEF}
    \zeta   = \frac{\alpha_0^2}{1 + \alpha_0^2} \,,\quad 
    \lambda = \frac{\beta_0}{2(1 + \alpha_0^2)} \,.
\end{equation}
The tight constraints on $\zeta$ from the Cassini mission imply that $\alpha_0^2 \lesssim 10^{-5}$. Furthermore, from equation~(\ref{eq:etaBW}) one then finds for the Nordtvedt parameter
\begin{equation}
    \eta \simeq 2\alpha_0^2(1 + \beta_0) \,.
\end{equation}
The sensitivity $s_a$ of a weakly self-gravitating body, like the WD companions to J0337+1715, can be calculated according to equation~(\ref{eq:sa_weak_field}):
\begin{equation}
    s_a \simeq -(1 + \beta_0) \, \varepsilon_{{\rm grav}, a} \,.
\end{equation}
For neutron stars, the absolute value of the sensitivity can become very large if $\beta_0 \lesssim -4.5$, a fact first discovered within $T_1(\alpha_0,\beta_0)$ gravity theories by \cite{DEF_1993}, and generally referred to as `spontaneous scalarisation'. As a result, even for arbitrarily small $\alpha_0$, the quantity $\alpha_0 s_a \simeq \zeta^{1/2} s_a$ remains at order unity.\footnote{The effective scalar coupling $\alpha_A$ used by \cite{DEF_1993} is linked to the sensitivity as defined here via $\alpha_A = \alpha_0(1 - 2s_A)$. Furthermore, $\alpha_0 \simeq \zeta^{1/2}$ for small $\alpha_0$.}

\begin{figure}[htb]
    \centering
    \includegraphics[width=\columnwidth]{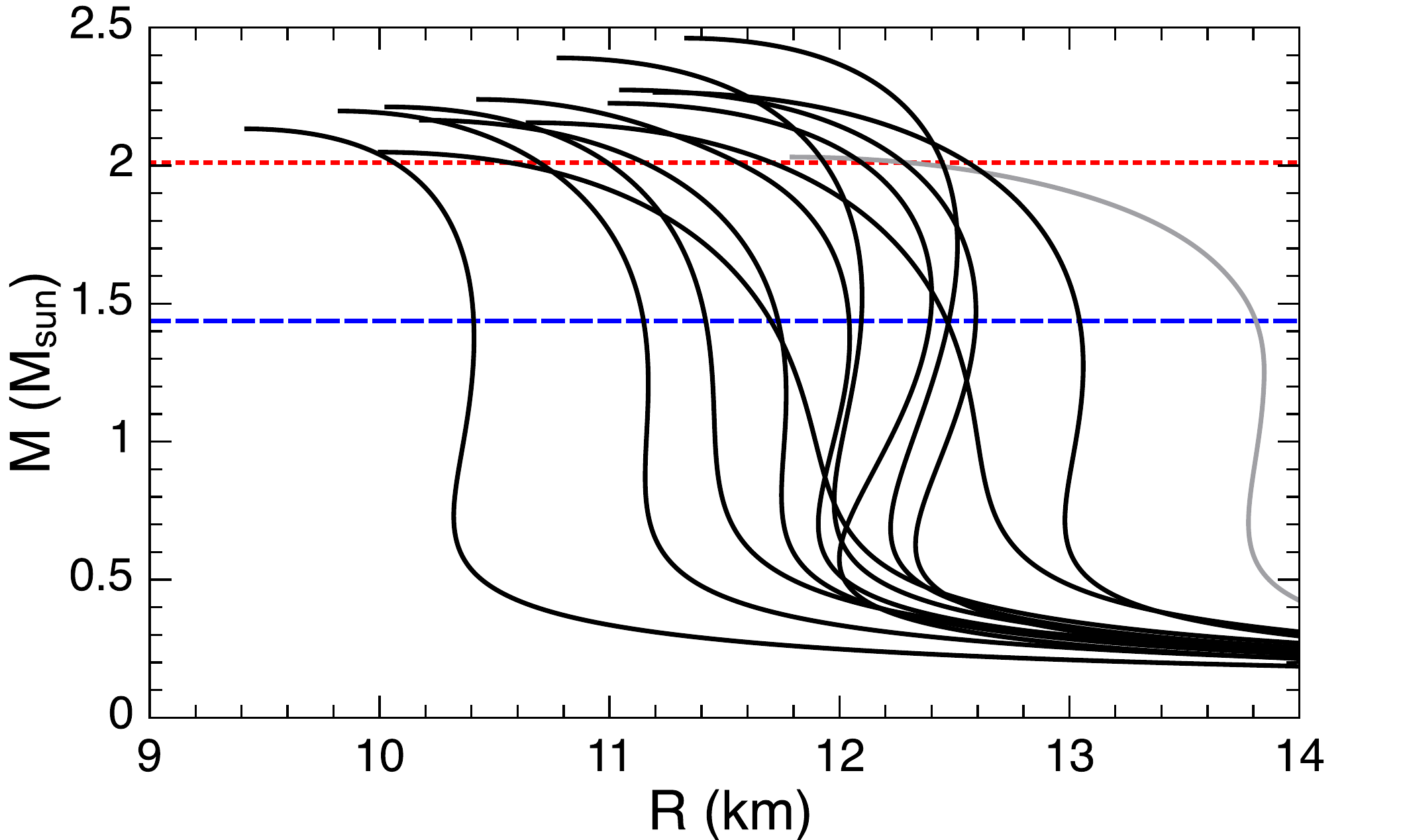}
    \caption{Radius-mass diagram for the 12 EOSs used in this paper to accomplish a good coverage of the range from soft to stiff EOSs, while still being in agreement with the tidal deformability test in the GW170817 LIGO/Virgo event \citep{PhysRevLett.119.161101,2018PhRvL.121p1101A} (black curves). The blue dashed line indicates the mass of PSR~J0337+1715, and the red dotted line corresponds to the most massive neutron star used in our combined test: PSR J0348+0432 \citep{antoniadis_massive_2013}. The black curves correspond to the following EOSs (from left to right in their intersection with the red dotted line): WFF1, SLy4, WFF2, AP4, BSk20, ENG, SLy9, AP3, BSk25, BSk21, MPA1, BSk22 (see \cite{Lattimer_2001} and https://compose.obspm.fr). Our stiffest EOS, BSk22, is also in agreement with the (more model dependent) constraint of $M_{\rm max} \lesssim 2.3\,M_\odot$ by \cite{Rezzolla_2018,Shibata_2019}. We have included EOS H4 (grey curve), which is disfavoured by GW170817, and therefore has not been used in Figs.~\ref{fig:a0b0-0337} and \ref{fig:a0b0}. All these EOSs also agree with the latest constraints from NICER \citep{Miller_2019}.}
    \label{fig:EOSs}
\end{figure}

A special case of $T_1(\alpha_0,\beta_0)$ is JFBD gravity, for which $\beta_0 = 0$. In that case $\lambda = 0$, and the coupling function is a constant:
\begin{equation}\label{eq:omegaBD}
    \omega(\phi) =\frac{1}{2\alpha_0^2} - \frac{3}{2} 
                 = \frac{1}{2\zeta} - 2 \equiv \omega_{\rm BD} > -\frac{3}{2}\,,
\end{equation}
which is called the Brans-Dicke parameter. For $\omega_{\rm BD} \rightarrow \infty$, that is,\ $\alpha_0$ and $\zeta \rightarrow 0$, JFBD gravity approaches GR. We obtain the most conservative limits on JFBD when using the stiffest EOS from our set of viable EOSs (see Fig.~\ref{fig:EOSs}), that is,\ BSk22. For this EOS, in JFBD gravity, the sensitivity of PSR~J0337+1715 has the value $s_{\rm p} = 0.149$. Most importantly, for $\zeta \lesssim 10^{2}$, this value is practically independent of $\zeta$ \citep{Shibata_2014,Shao_2017}. Hence, equation~(\ref{eq:zsplimit}) can directly be converted into limits on the coupling parameter:
\begin{equation}
    \zeta = (-1 \pm 6) \times 10^{-6} \quad \mbox{(95\% C.L.)} \,.
\end{equation}
Consequently, using equation (\ref{eq:omegaBD}), while keeping in mind that according to equation (\ref{eq:zlDEF}) $\zeta \ge 0$, one finds
\begin{equation}
    \omega_{\rm BD} > 140\,000 \quad \mbox{(95\% C.L.)} \,.
\end{equation}
This limit is more than a factor of three larger, that is,\ more constraining, than the Cassini limit \citep{Will_book}. When using EOS H4, which is already disfavoured by the GW170817 LIGO/Virgo event, we find $\omega_{\rm BD} > 130\,000$, which is only marginally weaker than the above limit. Just to illustrate the EOS dependence of the limit on JFBD gravity, for the soft EOS WFF1 (outer/left in Fig.~\ref{fig:EOSs}), the lower limit for $\omega_{\rm BD}$ increases to 180\,000.

\begin{figure}[htb]
    \centering
    \includegraphics[width=\columnwidth]{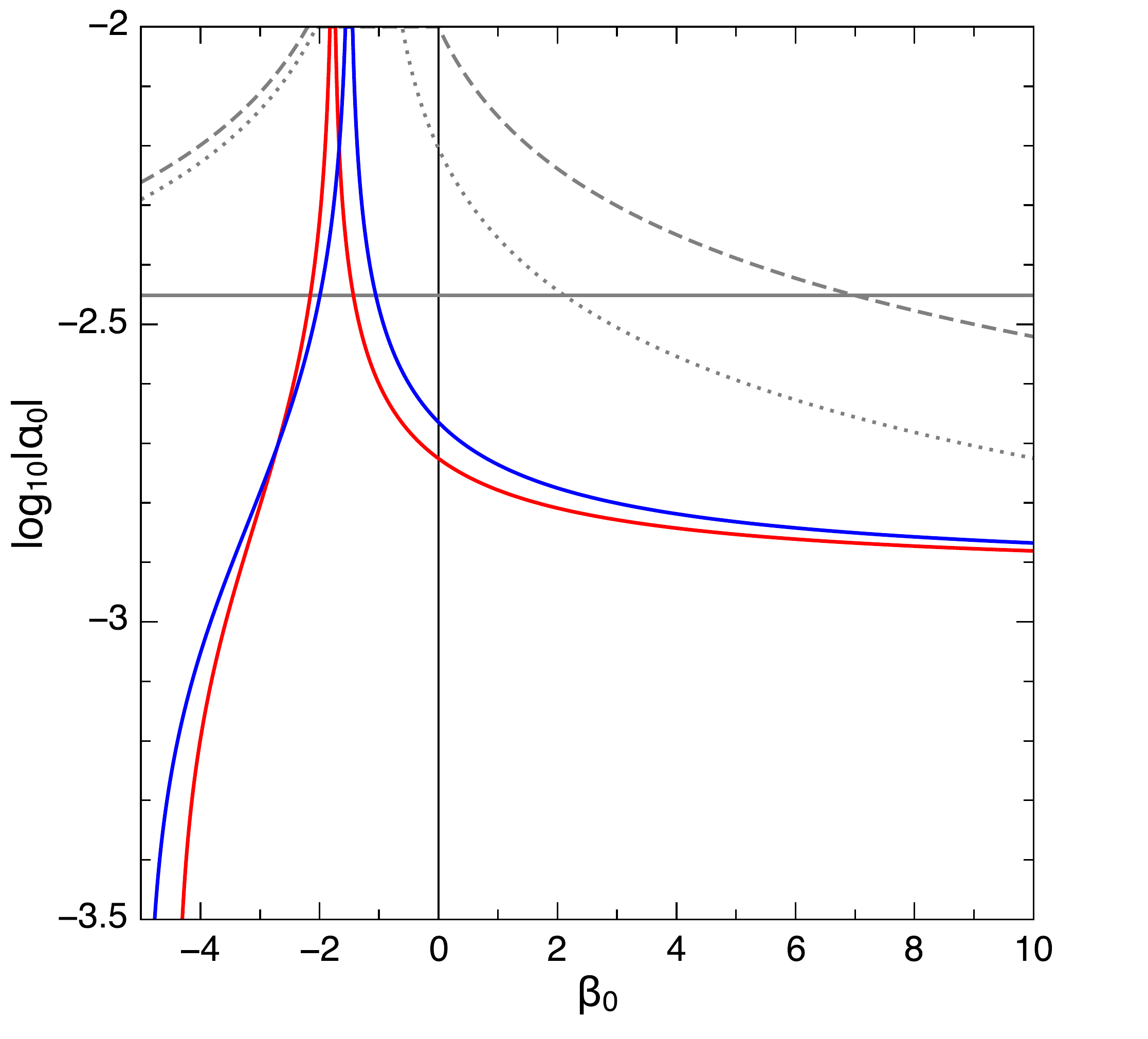}
    \caption{PSR J0337+1715 constraints on the $\alpha_0$-$\beta_0$ space of scalar-tensor theories \citep{DEF_1993,damour_tensor-scalar_1996} from equation~(\ref{eq:limit}): the area under the curve is still allowed by experiments. Two different neutron-star equations of state are used: a soft one, WFF1 (red curve), and a stiff one, BSk22 (blue curve). The two solid lines use the SEP constraint of this paper.  The grey curves show the 2$\sigma$-limits from Solar-system experiment: Cassini (solid), LLR (dashed), MESSENGER (dotted). JFBD gravity corresponds to $\beta_0 = 0$ (thin vertical line).}
    \label{fig:a0b0-0337}
\end{figure}

\begin{figure}[htb]
    \centering
    \includegraphics[width=\columnwidth]{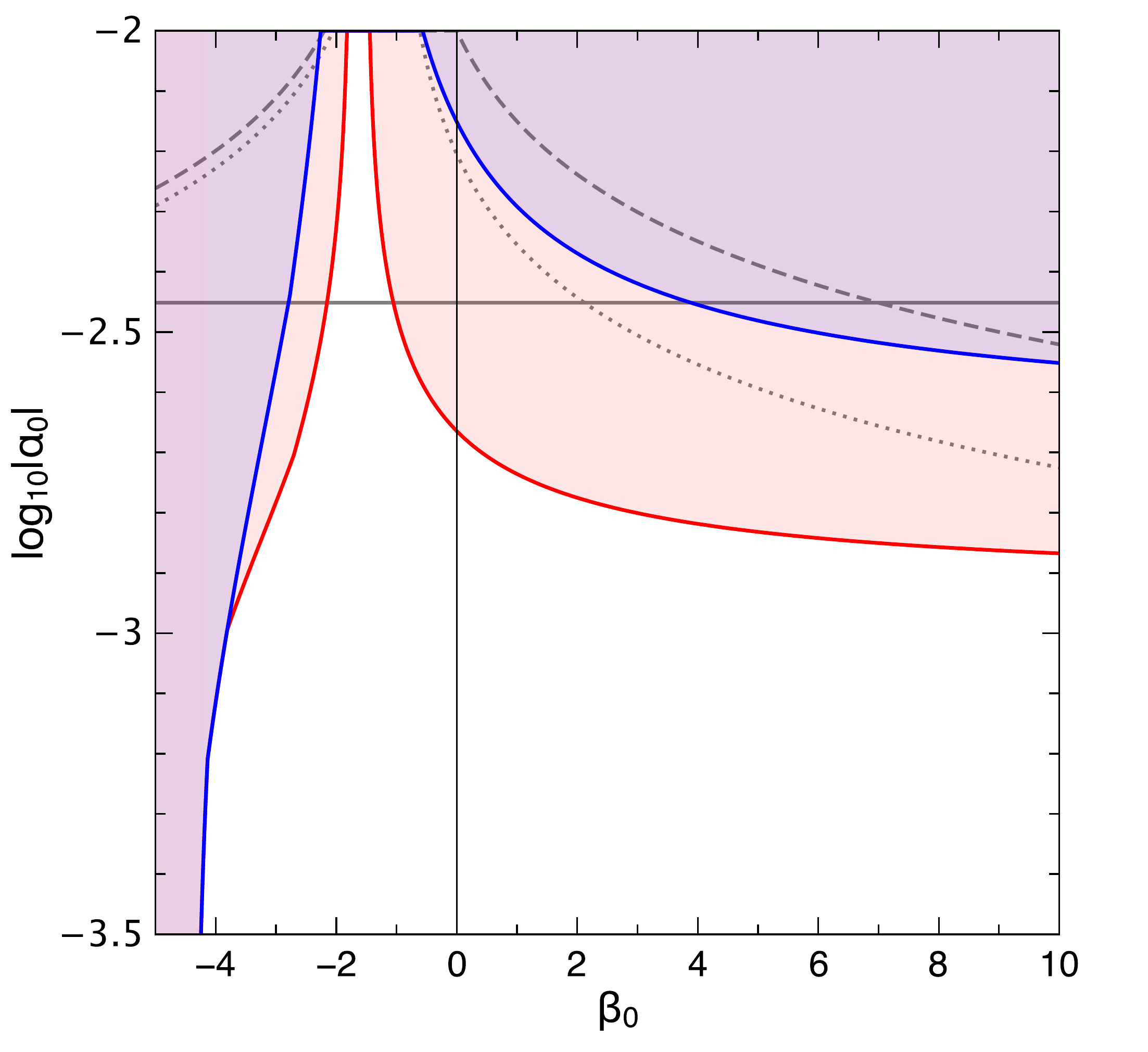}
    \caption{Combined EOS-agnostic pulsar constraints on the $\alpha_0$-$\beta_0$ space of scalar-tensor theories \citep{DEF_1993,damour_tensor-scalar_1996} from equation~(\ref{eq:limit}): the area under the curve is still allowed by experiments. The blue curve is the result of a combination of dipolar radiation tests from six pulsar-WD binary systems (see text for details). The red curve indicates the improvement when the constraints from this paper are added to the dipolar radiation tests. The grey curves show the 2$\sigma$-limits from Solar-system experiment: Cassini (solid), LLR (dashed), MESSENGER (dotted). JFBD gravity corresponds to $\beta_0 = 0$ (thin vertical line).}
    \label{fig:a0b0}
\end{figure}

While a stiffer EOS gives a more conservative limit for JFBD gravity, such a general statement is no longer true for the whole $\alpha_0$-$\beta_0$ parameter space of $T_1(\alpha_0,\beta_0)$. In particular for certain negative values of $\beta_0$, a softer EOS can be more conservative. For low and medium mass neutron stars, like PSR~J0337+1715, the $\beta_0$ range where that is the case is rather small (see Fig.~\ref{fig:a0b0-0337}). For high mass neutron stars the situation is quite different. A soft EOS that has a maximum mass close to the mass of the neutron star leads to considerably weaker limits for all $\beta_0 \lesssim -2$ \citep{Shibata_2014,Shao_2017}. This is of particular importance for constraints from pulsars like PSR~J0348+0432 \citep{antoniadis_massive_2013} (see Fig.~\ref{fig:EOSs}). Hence for our combined constraints on the parameter space of $T_1(\alpha_0,\beta_0)$ theories we used a set of EOSs that provide a good coverage of the range from soft to stiff. Furthermore, if for a given point $(\alpha_0,\beta_0)$, which corresponds to a specific gravity theory, there is a single EOS from our set with which all pulsar constraints are fulfilled then this point in the theory space is not excluded. For our joint analysis we have used the UFF results from this paper in combination with the dipolar radiation tests of PSRs J1012+5307 \citep{desvignes_2016,antoniadis_2016}, J1141$-$6545 \citep{bhat_2008}, J1738+0333 \citep{freire_1738+0333,Zhu_1713}, J1909$-$3744 \citep{desvignes_2016,Arzoumanian_2018}, and J2222$-$0137 \citep{cognard_2017}. Our results are shown in Fig.~\ref{fig:a0b0}. 

\section{Conclusions}
\label{sec:summary}

We described in this paper a test of the universality of free fall (UFF) with the pulsar in a triple star system, PSR~J0337+1715. The result we obtain for the UFF violation parameter for the MSP is $\Delta = (+0.5 \pm 1.8) \, \times\, 10^{-6}$ (95\% C.L.), which can be stated as a limit, $ | \Delta | < 2.05 \, \times\, 10^{-6}$ (also 95\,\% C.L.). This represents 30\% improvement over the previous test using the same pulsar \citep{archibald_universality_2018}. Interestingly, although we obtain a similar value for $\Delta$, the nature of our limit is different: the uncertainty reported in this work is statistical, while the result of \citet{archibald_universality_2018} is largely made of a systematic uncertainty which we did not find necessary in the present analysis. This particular difference makes our two limits difficult to compare in absence of a physically motivated model for the systematic bias, but should also provide an independent verification of the solidity of the result.

Furthermore, in a generic approach we also provide limits for three post-Newtonian strong-field parameters of the three-body interaction, and discuss in detail the relevance of these limits. In view of other binary pulsar limits, it seems that these limits might be of interest only in very specific situations.

As for \cite{archibald_universality_2018}, our results are fully consistent with the predictions of GR. This limit strongly constrains SEP violation and any alternative theories of gravity that predict a violation of the universality of free fall for self-gravitating masses (GWEP), particularly for neutron stars with masses similar to that of PSR~J0337+1715. In this paper we explicitly calculate these constraints for a wide class of gravity theories, and as a part of this derive EOS-independent constraints on the parameter space of quadratic mono-scalar-tensor gravity. Specifically, for the coupling parameter of Jordan-Fierz-Brans-Dicke (JFBD) gravity we find $\omega_{\rm BD} > 140\,000$, which is the so far the tightest limit for this scalar-tensor theory. We also present new constraints for Damour--Esposito-Far\`ese (DEF) gravity, a quadratic extension of JFBD gravity. We combine our limit with limits from binary pulsar experiments while accounting for uncertainties in our knowledge of the equation of state (EOS) of neutron-star matter.

In what remains of the paper, we make a detailed comparison of this experiment with the best previous constraints on GWEP/SEP violation. In Section \ref{sec:comparison} we make a more detailed comparison with the experiment by \cite{archibald_universality_2018}. In Section~\ref{sec:radiative}, we compare our experiment to radiative experiments from binary pulsars, which have also produced strong and complementary constraints on GWEP violation via their strong constraints on the emission of dipolar gravitational waves.  Furthermore, we compare the present limit with potential future limits on dipolar radiation from binary neutron-star and neutron star-black hole mergers.

In all of these experiments, no GWEP violation can be detected; gravity behaves, to within observable precision, as described by GR, which is conjectured to be the only viable theory which fully embodies the SEP.

\subsection{Comparison with previous work on J0337+1715}
\label{sec:comparison}

This work distinguishes itself from the \citet{archibald_universality_2018} on the following points: 
\begin{enumerate}
    \item Independent data set;
    \item Independent timing model including additional effects;
    \item Statistics-limited versus systematics-limited accuracy;
    \item Tension in the mass measurements and the first measurement of $\Omega_{\rm O}$;
    \item Generic test of those strong-field post-Newtonian parameters, which are a-priori unconstrained, even within a broad class of scalar-tensor theories.
    \item EOS-agnostic constraints on DEF gravity, while accounting for the latest observational constraints on the range of EOSs.
\end{enumerate}
Point 1) benefits from the well sampled timing data acquired by the Nan\c cay radio telescope alone (Section \ref{sec:Observation}). All the observations used here were conducted within the same frequency range (1.2-1.7GHz) and with the same environmental conditions since Nan\c cay is a meridian Kraus design telescope. The NUPPI instrumentation is also routinely used for long term high-precision timing providing excellent and stable results. The instrumentation did not change since its installation in 2011 and there is no need for any time jump in the whole dataset since 2011.

Point 2) makes use of \textsc{nutimo} (Section \ref{sec:numerical_model}) which has the specificity of allowing for a fully self-consistent treatment of astrometric parameters via the binding with \textsc{tempo2} and the inclusion of Kopeikin and Shklovskii delays in the model. The model in \citet{archibald_universality_2018} did not include these delays and used a local linear approximation for astrometric corrections. The argument in favour of such proxy was that any systematic effect caused by these approximations should not affect the main SEP signature which has a different frequency. However, we observe that the astrometry then found differs significantly from prior knowledge and in particular Gaia observations which led \citet{archibald_universality_2018} to acknowledge that the resulting astrometry should not be used for other applications. In addition, \textsc{nutimo} also fits consistently for DM and DM variations and allows to check for local epoch DM variations (DMX parameters in \textsc{tempo2}) which revealed no fluctuations over time. Note that \citet{archibald_universality_2018} did fit DM over 1 year time intervals and marginalised over these parameters using the solution of a least-square fit. We also included the aberration delay that neither \citet{ransom_millisecond_2014} nor \citet{archibald_universality_2018} mention. This delay has a very small amplitude (sum of $\sim30\,\si{ns}$ sinusoid at the outer period and a $\sim 0.1\,\si{\mu s}$ sinusoid at the inner period) and therefore can easily be absorbed by other parameter in a fit, but still creates a signal of magnitude larger than the expected SEP sensitivity. In \textsc{nutimo}, potential systematic effects that may not be accounted for by the model are absorbed in a re-scaling of the error bars of the times of arrival via the EFAC parameter which ensures a reduced $\chi^2$ equal to unity. This in turns conservatively increases uncertainties on the posterior parameters. 

Point 3) arises from the fact that, in \citet{archibald_universality_2018}, most of the total reported uncertainty of $\Delta$ ($\pm 0.74\times 10^{-6}$ at 68\% CL) is associated with systematic uncertainties while in this work our uncertainty is mostly statistical. We account for unmodelled systematic effects, mostly a red-noise component, via the EFAC parameter which is responsible for a modest and conservative widening of $\sim 8\%$ of the uncertainties. The statistical uncertainty of \citet{archibald_universality_2018} is estimated using MCMC sampling similarly as we do in this work and results in $\Delta = (-1.1 \pm 0.2) \times 10^{-6}$ (68\% CL). Taken alone, this would signify a 5-sigma SEP violation. However, the authors argue that most of the uncertainty comes from unaccounted systematic effects which could generate a signal at the signature frequency of an SEP violation. In other words, it is claimed that the accuracy is systematics-limited while in this work we are statistics-limited. The physical mechanism of the systematics being unknown, \citet{archibald_universality_2018} propose to model systematics using an empirical stochastic model where the extraneous signal is a weighted sum of sine and cosine functions at frequencies $kf_{\rm i} + lf_{\rm o}$ ($k,l$ being small integers) whose weights are drawn from a single Gaussian distribution for each particular realisation. In order to sample the distribution of $\Delta$ caused by different realisations of the (stochastic) systematics,  \citet{archibald_universality_2018} bootstrapped many sets of synthetic data from the model, re-fitted the orbital model, and thus obtained a value of $\Delta$ for each synthetic dataset. This estimate heavily depends on the modelling choices for which no physical justification is currently available. It also seems unlikely that systematics should occur at frequencies $kf_{\rm i} + lf_{\rm o}$ if the physical mechanism is unrelated to orbital motion (as an SEP violation would be), unless the signal at these frequencies is the tail (in Fourier space) of a systematic signal which peaks at a different frequency, but should then be seen in a periodogram such as Figure \ref{fig:periodogram}. In this respect, Figure \ref{fig:periodogram} does not suggest that we should consider a systematics-limited regime in the present work. Although some systematics are present as red noise (see Section \ref{sec:residualanalysis}), there is no sign of an additional signal around the signature frequency.

Point 4) is about comparing the parameter set of Table \ref{tab:fitresults} with the results of \citet{archibald_universality_2018}. A direct comparison is not straightforward because of i) slightly different definition of the parameters (see section \ref{sec:param}) and ii) the fact that most of the parameters are not constants of motion but are defined either at the reference time $T_\mathrm{ref}$ or $T_\mathrm{pos}$. To minimise the span of numerical computations we do not use the same reference time as \citet{archibald_universality_2018}. However, one can compare masses which are constants of motion (and, in the same way, $\Delta$). The values reported in \citet{archibald_universality_2018}, $m_{\rm p} = 1.4359(3)\Msol, m_{\rm i} = 0.19730(4)\Msol, m_{\rm o} = 0.40962(9)\Msol$\footnote{The number between brackets gives the uncertainty on the last digit(s).}, whose statistical 68\% confidence intervals are about 5 times better than ours - similar to the ratio between the uncertainties on $\Delta$ -
are in tension with the values we report in Table \ref{tab:fitresults}, with $\Delta m_{\rm p} \simeq 2.9\sigma, \Delta m_{\rm i} \simeq 2.6\sigma, \Delta m_{\rm o} \simeq 2.4\sigma$ where $\sigma$ is half of the 68\% confidence interval reported in this paper, these differences are much more significant than for $\Delta$. Due to the very large correlation between $\Delta$ and the orbital parameters on which depend the masses (period and semi-major axis), the systematic uncertainty estimated in \citet{archibald_universality_2018} for $\Delta$ should be similar for the masses (but not reported) and partly release the tension.

In addition, we report the first measurement of the outer longitude of ascending node, $\Omega_{\rm O}$, which was deemed unconstrained in \citet{archibald_universality_2018} although the dispersion of the fit residuals in \citet{archibald_universality_2018} is smaller than ours. We speculate that the absence of Kopeikin delay in their analysis prevented that measurement. However, this parameter is uncorrelated with $\Delta$ and should therefore not affect the SEP test, but its absence should bias astrometric parameters.

Point 5) is related to  our implementation of the first post-Newtonian equations of motion, derived from the modified Einstein-Infeld-Hoffmann Lagrangian for strongly self-gravitating masses (Appendix \ref{sec:eqofmotion}). In this we use two different approaches, a (mostly) generic one where three of the 12 1PN strong-field parameters are unconstrained by Solar System experiments, and a second approach where, under additional assumptions, all the strong-field 1PN parameters are tightly constrained by adopting binary pulsar constraints for the neutron star sensitivity and its derivative. A detailed motivation for the two different approaches is given in Section~\ref{sec:tests_theory}. In the first approach we find generic limits for the remaining three 1PN strong-field parameters, which however are not very tight, and therefore generally not of particular interest. \cite{archibald_universality_2018} do not provide an equally generic analysis as done in our first approach.

Point 6) refers to the constraints of quadratic mono-scalar-tensor gravity, where \cite{archibald_universality_2018} have used a single (outdated) EOS. In our combined tests we have fully accounted for our imperfect knowledge of the EOS of neutron-star matter, and used a set of modern EOSs that covers the range from soft to stiff EOSs. A reason for that is the fact that the most conservative pulsar limits do not always come from the stiffest EOS. Our set of EOSs is in agreement with the latest constraints from LIGO/Virgo and NICER, and can account for the largest neutron star masses measured to date \citep{antoniadis_massive_2013, Cromartie_2019}. We would like to point out, that the recent limits of \cite{capano2019gw170817} exclude some of the stiffer EOSs used in our analysis, which consequently leads to even more stringent constraints than the ones shown in Fig.~\ref{fig:a0b0}, in particular for $\beta_0 \gtrsim -1$. In view of this, our limits can be considered as conservative.

\subsection{Comparison with radiative tests}
\label{sec:radiative}

In GR, the lowest source multipole moment that generates gravitational waves is the quadrupole moment \citep{Thorne_1980}. In alternatives to GR, however, one finds lower multipoles, where for the dynamics of a binary system, the dipole moment is the most important one. The occurrence of these lower multipoles is closely related to a violation of the SEP \citep[see e.g.\ ][for a discussion]{Will_book}. In scalar-tensor theories, for instance, an asymmetry in the sensitivities $s_a$ in a binary system gives rise to scalar dipolar radiation (see equation~(\ref{eq:DGW})). While a difference in sensitivity is also the reason for a violation of GWEP, where masses with different compactness are falling differently in an external gravitational field  (see equation~(\ref{eq:Gsasb})). In a sense, the UFF experiment with PSR~J0337+1715 and constraints on dipolar radiation damping with binary pulsars are exploring two different sides of the same coin. The limits in Fig.~\ref{fig:a0b0} show that currently for a large part of the parameter space, the test with PSR~J0337+1715 is more constraining than dipolar radiation tests from binary pulsars. For sufficiently, negative $\beta_0$, however, gravitational wave tests with binary pulsars become more constraining, in particular for small $\alpha_0$. We have a more detailed discussion on this further below.  

\begin{figure}[htb]
    \centering
    \includegraphics[width=\columnwidth]{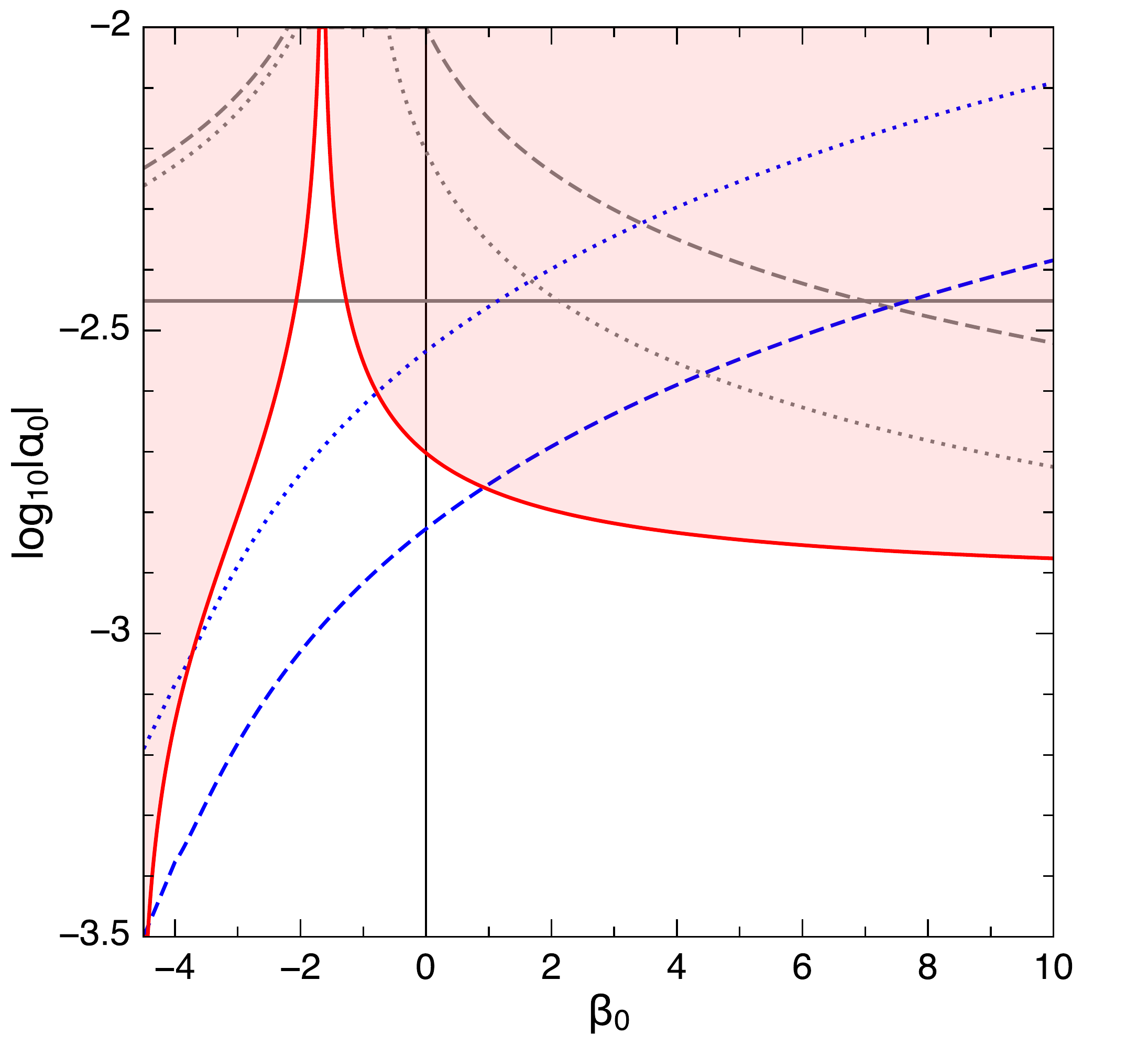}
    \caption{Comparison of the PSR~J0337+1715 constraints of this paper with expected constraints from future gravitational wave observatories of a single binary neutron-star merger: Cosmic Explorer (CE; blue dotted curve) and Einstein Telescope (ET; blue dashed curve). CE and ET curves are taken from Fig.~9 in \cite{Shao_2017}. Like for the CE and ET curves, the PSR~J0337+1715 curve is also based on EOS AP4. Solar System constraints (grey) are as in Fig.~\ref{fig:a0b0-0337}. JFBD gravity corresponds to $\beta_0 = 0$ (thin vertical line).}
    \label{fig:J0337-CE-ET}
\end{figure}

Gravitational wave observation of a double neutron-star merger can also be used to constrain the emission of dipolar gravitational waves, as has been done for the first LIGO/Virgo binary neutron-star merger GW170817 \citep{GW170817-GR-test}. Limits on scalar-tensor theories as discussed here, from LIGO/Virgo observations, however, are not expected to be competitive with Solar System and pulsar experiments for most of the parameter space \citep[see][]{Shao_2017}. Future ground based gravitational wave detectors have the potential to improve on limits presented here, in particular in a $\beta_0$ range which is difficult to constrain with pulsar experiments (see Fig.~\ref{fig:J0337-CE-ET}). Future gravitational wave observations of mixed (black hole + neutron star) mergers, in particular the combination of multiple events or the combination of ground and space based gravitational-wave observatories promise significant improvements \citep[see e.g.][]{Carson_2019}.

As a final comment, there is an important difference between the UFF test conducted with PSR~J0337+1715 and dipolar radiation tests. As discussed in the previous subsection, in the regime of spontaneous scalarisation, the neutron star charge can become (almost) independent of the parameter $\zeta$ in the sense that $\zeta s_{\rm p}^2 \sim 1$ remains practically fixed for $\zeta \rightarrow 0$. In such a situation, the effective gravitational constant in the interaction between a neutron star and a white dwarf becomes indistinguishable from $G_{\rm N}$ and the test with J0337+1715 becomes practically insensitive to such deviations from GR. Dipolar radiation test with pulsar-white dwarf systems, in contrast, are extremely constraining with respect to such scalarisation phenomena, as can be seen from equation~(\ref{eq:DGW}). More generally, in situations where only the strong field of a neutron star can source additional (long-range) gravitational fields that lead to deviations from GR, the UFF test with PSR~J0337+1715 cannot place any constraints, in contrast to radiative tests. Hence, both types of tests are complementary and valuable. Binary pulsar tests have already tightly constrained the occurrence of spontaneous scalarisation in neutron stars. However, depending on the EOS and mass of the neutron star, spontaneous scalarisation is not yet fully ruled out by such experiments \citep{Shao_2017}.


\begin{appendix}

\section{Strong-field equations of motion \label{sec:eqofmotion}}

In our timing model, the motion of the three bodies follows the equations of motion derived from the post-Galiean-invariant $n$-body Lagrangian of the modified Einstein-Infeld-Hoffmann (mEIH) formalism \citep{will_theory_1993, damour_strong-field_1992}. The mEIH equations of motion describe the first post-Newtonian dynamics of a $n$-body system which also contains strongly self-gravitating masses, under the assumption that the gravitational interaction is Poincar\'e invariant. Furthermore, it assumes that there are no `asymmetric' terms in the Lagrangian, which are anyway absent in many well motivated theories of gravity \citep[see the discussion in][]{Nordtvedt_1985, damour_strong-field_1992}. The mEIH formalsim is a generalisation of the parametrised post-Newtonian (PPN) equations of motion for fully conservative theories with $\xi = 0$, in order to include effects related to the strong internal fields of strongly self-gravitating objects, like neutron stars. The mEIH Lagrangian can be written as \citep[cf.][]{DEF_1992}
\begin{eqnarray}
\mathscr{L} & = & \sum_{a=1}^n \left(- m_a c^2 +  m_a \frac{ v_a^2 }{2} +  m_a\frac{v_a^4}{8 c^2} \right) 
\nonumber\\ &&
  + \frac{1}{2} \sum_{a=1}^n \sum _{b\neq a}^n  \left\{ \frac{ G_{ab} m_a m_b }{ r_{ab} } 
  \left[1 -\frac{(\vec{v}_a \cdot \vec{n}_{ab})(\vec{v}_b \cdot \vec{n}_{ab})}{ 2 c^2} 
\right. \right. \nonumber \\ && \qquad \left.\left.
- \frac{7}{2} \frac{ \vec{v}_a\cdot \vec{v}_b }{c^2} 
   + \frac{3}{2} \left(\frac{ v_a^2 }{c^2 } + \frac{v_b^2 }{c^2}\right)
   + \bar{\gamma}_{ab}\frac{\left( \vec{v}_a - \vec{v}_b\right)^2}{c^2} \right]
\right. \nonumber \\ && \qquad \left.
   -  \sum _{c\neq a}^n \frac{G_{ab}G_{ac} m_a m_b m_c }{c^2 r_{ab} r_{ac} } (1 + 2\bar{\beta}_{bc}^a) \right\},
\label{eqtr:lagrangian1pn}
\end{eqnarray}
where $m_a$ are the inertial masses with coordinate positions $\vx_a$ and coordinate velocities $\vv_a$, $r_{ab} = \|\vx_a - \vx_b\|$, $v_a = \|\vv_a\|$, and $\vn_{ab} = (\vx_b - \vx_a)/r_{ab}$. The quantities $\bar{\gamma}_{ab} =  \gamma_{ab} - 1$, $\bar{\beta}_{bc}^a = \beta_{bc}^a - 1$ and $G_{ab}$ are the effective strong-field interaction constants. The unbarred quantities are the strong-field generalisation of the PPN parameters $\gamma_{\rm PPN}$ and $\beta_{\rm PPN}$ (Eddington parameters). The strong-field parameters satisfy the symmetries $G_{ab} = G_{ba}$ ($a \ne b$), $\bar{\gamma}_{ab} = \bar{\gamma}_{ba}$ ($a \ne b$), and $\bar{\beta}_{bc}^a = \bar{\beta}_{cb}^a$ ($a \ne b$, $a \ne c$). The body-dependent effective strong-field interaction constants depend on the details of the underlying gravity theory as well as the structure of the individual bodies. Hence, in the most general case of a three-body system one has three different effective gravitational constants $G_{ab}$, three different $\bar\gamma_{ab}$, and nine different $\bar{\beta}_{cb}^a$. In GR, due to the fulfilment of SEP and the corresponding effacement of the internal structure \citep[see e.g.][]{Damour_1987}, one has $G_{ab} = G_{\rm N}$ and $\bar{\gamma}_{ab} = \bar{\beta}_{bc}^a = 0$.

One can then use the Euler-Lagrange equations \citep{will_theory_1993} to derive the equations of motion for each body: 
\begin{eqnarray}
\label{eqtr:motion}
\ddot{\vec{x}}_a & = & \sum_{b\neq a} \frac{ G_{ab} m_b}{r_{ab}^2}  \vn_{ab} \left[   1 - \frac{1}{c^2}\left(4 \vv_a\cdot \vv_b - \vv_a^2 - 2 \vv_b^2 \right.\right. \nonumber\\
& & \left.\left. + \frac{3}{2}(\vv_b \cdot \vn_{ab})^2 - \bar{\gamma}_{ab} \left(\vec{v}_a - \vec{v}_b\right)^2  \right)\right] \nonumber \\
&  & + \sum_{b\neq a} \frac{ G_{ab} m_b  }{r_{ab}^2 c^2}\left(\vv_b -\vv_a \right) \left[  \vn_{ab} \cdot \left(4\vv_a - 3\vv_b - 2\bar{\gamma}_{ab}(\vv_b - \vv_a)\right) \right]   \nonumber \\ 
& & +  \sum_{b\neq a}\sum_{c \neq b}  \frac{ G_{ab}G_{bc} m_bm_c}{r_{ab}r_{bc} c^2} \left[\frac{1}{r_{bc}}\left(\frac{1}{2}(\vn_{ab}\cdot \vn_{bc})\vn_{ab} + \frac{7}{2}\vn_{bc}\right)  \right. \nonumber \\ 
& & \left. - \frac{\vn_{ab}}{r_{ab}} +  2\bar{\gamma}_{ab}\frac{\vn_{bc}}{r_{bc}} - 2\bar{\beta}_{ca}^b \frac{\vn_{ab} }{r_{ab}} \right]    \nonumber \\
& & -\sum_{b\neq a}\sum_{c \neq a}  \frac{ G_{ab}G_{ac} m_b m_c}{r_{ab}^2 r_{ac} c^2} \vn_{ab} \left[4 +  2\bar{\gamma}_{ac} + 2\bar{\beta}_{bc}^a \right]  \,.
\end{eqnarray}
%
In the weak-field limit one can check that this equation does give the PPN equation of motion (e.g. \citealt{soffel_relativity_1989, will_theory_1993}). 

Conserved quantities are key elements to check the numerical implementation and accuracy of the equations of motion. We have used the Hamiltonian (conservation of energy), and the momentum and position of the centre of mass of the system. The last two are also necessary to derive the initial conditions of the system.

The Hamiltonian corresponding to equation \eqref{eqtr:lagrangian1pn} is derived using the Legendre transform $ \mathscr{H} = \sum_a \vv_a \cdot\frac{\partial{\mathscr{L}}}{\partial\vv_a} - \mathscr{L}$,
\begin{eqnarray}
\label{eqtr:hamiltonian1pn}
\mathscr{H}  &=  & \sum_a \left\{ m_a c^2 + \frac{1}{2} m_a v_a^2 + \frac{3}{8}m_a \frac{v_a^4}{c^2}  \right. \nonumber \\
 & &  +\frac{1}{2}\sum_{b \neq a} \left[ -\frac{G_{ab}m_a m_b }{r_{ab}} \left(1 + \frac{7}{2}\frac{\vv_{a}\cdot\vv_{b} }{c^2} \right. \right. \nonumber \\
 & & \left. + \frac{1}{2c^2} (\vv_a\cdot \vn_{ab})(\vv_b\cdot \vn_{ab}) - 3 \frac{v_a^2}{c^2} + \bar{\gamma}_{ab}\frac{\left(\vv_a - \vv_b\right)^2}{c^2} \right)  \nonumber \\
& &  \left. \left. + \frac{1}{2}\sum_{c \neq a} \left(1 + 2\bar{\beta}_{bc}^a\right)\frac{G_{ab} m_a m_b }{r_{ab}}\frac{G_{ac} m_c }{r_{ac} c^2} \right] \right\}.
\end{eqnarray}

The momentum of the centre of mass is given by the same expression as in GR only with the replacement $G \rightarrow G_{ab}$,  $\vec{P} = \sum_a \frac{\partial{\mathscr{L}}}{\partial \vv_a} $ and 
\begin{eqnarray}
\label{eqtr:momentum}
\vec{P} & = & \sum_a \left[ m_a \vv_a\left(1 + \frac{v_a^2}{2c^2} - \frac{1}{2} \sum_{b\neq a} \frac{G_{ab} m_b }{c^2 r_{ab}} \right) \right. \nonumber \\
 & & - \left. \frac{1}{2} \sum_{b\neq a} \frac{G_{ab} m_b }{c^2 r_{ab}}  (\vv_a \cdot \vn_{ab}) \vn_{ab}  \right].
\end{eqnarray}

The centre-of-mass position $\vec{X}$ satisfies $(\mathscr{H}/c^2)\frac{\mathrm{d}X}{\mathrm{d}t} = \vec{P}$ (see e.g. \citealt{will_incorporating_2014}), 
\begin{equation}
\label{eqtr:cdm1pn}
\frac{\mathscr{H}}{c^2} \vec{X} = \sum_a  m_a \vx_a\left(1 + \frac{v_a^2}{2c^2} - \frac{1}{2} \sum_{b\neq a} \frac{G_{ab} m_b }{c^2 r_{ab}}\right) + \mathscr{O}(c^{-4}).
\end{equation}

\end{appendix}

\begin{acknowledgements}
G. Voisin acknowledges support of the European Research Council, under the European Unions Horizon 2020 research and innovation programme (grant agreement No.715051; Spiders).
GV would like to thank F. Mottez and R. P. Breton for their support and helpful discussions during this project. GV also thanks A. Archibald for valuable discussions during the very early stages of this work.

\\
This work was granted access to the HPC resources of MesoPSL financed
by the Region Ile de France and the project Equip@Meso (reference
ANR-10-EQPX-29-01) of the programme Investissements d’Avenir supervised
by the Agence Nationale pour la Recherche. We acknowledge financial support from the Action Fédératrice PhyFOG funded by Paris Observatory and from the “Programme National Gravitation, Références, Astronomie, Métrologie (PNGRAM) funded by CNRS/INSU and CNES, 
France. GD, MK and NW gratefully acknowledge support from European Research Council (ERC) Synergy Grant `BlackHoleCam' Grant Agreement Number 610058.

\\
This work made use of the Scipy libraries (\url{www.scipy.org}).
\\
The authors would like to thank Lijing Shao for his valuable comments that helped to improve the manuscript.

\end{acknowledgements}

%
\bibliographystyle{aa} 
\bibliography{triple_system} 
%

\end{document}